\documentstyle[12pt]{article}
\begin{document}
\begin{titlepage}
\begin{center}
\vspace{4 cm}
{\Huge\bf Heavy neutrinos detection \\
in the Next Linear Collider} \\
\vspace{3 cm}
{\huge Ph.D.~Thesis} \\
\vspace{2 cm}
{\Large Janusz Gluza}
\vskip 1.5cm
Department of Field Theory and Particle Physics \\
University of Silesia\\
Katowice, Poland
\vspace{4 cm}
\end{center}
\ \\

January 1997.
\end{titlepage}
\pagenumbering{roman}
\
\begin{center}
Acknowledgments
\end{center}

\ \\

My special thanks are due to my supervisor Prof. Dr. Marek Zra\l ek 
for his constant
encouragement and interest. Our conversations were really
inspiring and helped me very much.

\ \\

I also appreciate the financial support from the Polish Committee for Scientific
Research, Grants No PB659/P03/95/08 and PB514/P03/96/11.

\ \\
\ \\
\ \\
\ \\

I'd like to dedicate this work to my Parents, my wife and children 
Marta and Marek, people who fill my life with happiness.
\newpage

\tableofcontents
\newpage

\pagenumbering{arabic}
\baselineskip 7 mm
\section{Introduction}
\setcounter{equation}{0}
\renewcommand{\theequation}{1.\arabic{equation}}

Since the beginning of the standard model's (SM's) construction
based on the $SU(3)_C \otimes SU(2)_L \otimes U(1)_Y$ gauge group 
\cite{sm} people realized it had
drawbacks\footnote{S. Glashow, one of the founders of this
model, said at his Nobel lecture: `Let me stress that I do not believe
that the standard theory will long survive as a correct and complete picture 
of physics' \cite{glash}.}. However, almost three decades after its birth 
(and two after Glashow's lecture) the SM seems to be in 
good shape. It appears that this model describes fundamental electroweak and
strong interactions very well and it is hard (till now impossible) to
find out something extraordinary, non-standard in nature\footnote{Maybe with two
exceptions. The $R_b$ parameter in LEPI at CERN,
is about 1.8 $\sigma$ above its SM's value (the situation warmly 
invited by advocators of theories with
supersymmetric extensions of the SM \cite{pok}) and the LSND
experiment \cite{lsnd} for which the most probable interpretation is that 
non-vanishing neutrino mass is for the first time observed there.}.
The experimental data has become incomparably better, but the theory has
remained essentially unmodified.

So, to find out anything beyond the SM, high-energy physicists go generally 
in one of  two
directions. First, they are continually improving experimental precision.
It's a common belief that with the shrinking
error in measurements, some experimental data can finally appear outside
its SM value. Second, they use higher and higher energies
in experiments. As an illustration of the first approach,
let's consider the LEPI program
where till the end of 1995 better and better precision of measurements
has been achieved. The pay-off for this was, however, unexpected. 
It appeared
that some `strange' phenomena from the macro-scale world started to be important
in par-excellence micro-world experiment as the LEPI is, too. Falling rains,
passing trains and the Moon's (and the Sun's) attraction 
(some people even joke that the LEPI produced the
first evidence for gravitational force unification) have influenced
and disturbed data collection \cite{moon}. 
So, most probably, in the future  higher energies
and new machines will be necessary if we are going to investigate fundamental 
interactions and make any progress in high-energy physics. However, even then,
extremely precise measurements from low-energy physics will be useful. For
instance,
in this Thesis interplay between these two seemingly
different approaches is undebated and all the most important conclusions
about high energy physics derived here are the result of  extraordinary
precision obtained in double beta or muon decays among other things.

Because such new experiments are sophisticated nowadays (and expensive),
it is necessary to investigate as many  phenomena as possible 
to find out, if these phenomena can be really revealed there.
This Thesis is about one such possible phenomenon in the
Next Linear Collider (NLC).
The NLC collider is one of the many names for the future $e^+e^-$ 
collider, a machine with a center of mass energy above 200 GeV (LEPII),
up to 1500-2000 GeV \cite{nlc},\cite{snow}.
The NLC will complement the other future accelerator, the high energy hadron collider
LHC at CERN. Together these two machines will provide a coherent
program for understanding the SM better and hopefully non-standard, new
physics as well.

Many new or un-established ideas such as anomaly couplings, additional 
(neutral and charged) gauge bosons, excited fermions and Higgs particles
can be tested in the NLC collider \cite{nlctests}. Heavy neutrinos are among 
these ideas
and I will deal with this case in this Thesis (for other than `NLC'
possibility of heavy neutrino detection - see
\cite{pr8},\cite{9_1}-\cite{9_7}).

Although there are three types of massless neutrinos in the SM,  almost all 
extensions
of the SM predict more than three massive neutrinos. The last LSND result seems
to have claimed, at least partly, this prediction.
We know so far that three known neutrinos are very light with masses
\cite{pdglight}
\begin{equation}
m_{\nu_e} \leq 15\;eV,\;m_{\nu_{\mu}} \leq 170\;keV,\;m_{\nu_{\tau}} \leq
24\;MeV.
\end{equation}
We know from the
negative search of new neutral states and from the measurement of Z decay width
at the LEPI too, that there are no neutrinos with a standard coupling to Z and mass
below $M_Z/2$ \cite{mz/2} or even below $M_Z$ if $BR(Z^0 \rightarrow \nu N) >
3 \cdot 10^{-5}$ \cite{bran}. The lack of detection of new neutrino states at the LEPI
indicates that if they exist, they will generally have large masses
($ \geq M_Z$), attainable in the NLC.

Such masses of heavy neutrinos $(100\;\mbox{\rm GeV} \leq M_N \leq
0(1)\;\mbox{\rm TeV})$ are natural for a theoretical model based
on the $SU(2)_L \otimes SU(2)_R \otimes U(1)_{B-L}$ gauge group \cite{lrm},
the so-called left-right symmetric model (LR). In this model left-handed and 
right-handed
(lepton and quark) fields are treated in the same way: there are 3 left-handed
and 3 right-handed doublets (one left-handed and right-handed doublet for
each generation). This symmetry causes that apart from three light and known
(left-handed) neutrinos we have 3 heavy $(M_N>M_Z)$ ones. Their masses are 
utmost at
the level at which the left-right symmetry is broken, e.g. of the order of the
right-handed gauge boson mass $M_{W_R}$. Up-to-date constraints gives
$M_{W_R} \geq $406 GeV \cite{wr} (652 GeV \cite{wrhad}), so we can expect to look 
for heavy neutrinos of the LR model in the NLC collider.

The second easy extension of the SM which includes heavy neutrino particles
is the SM supplemented by right-handed singlet neutrino fields (RHS model). 
Such fields
appear naturally in the SO(10) Grand Unified Theory (GUT) in which quarks and
leptons are accommodated into one fundamental 16-dimensional representation
(3 colours of left- and right-handed u and d quarks plus $e_L$ plus $e_R$
plus $\nu_L$ plus $N_R$). In this GUT model, the natural mass of such right-handed
heavy neutrinos is of the order of the GUT scale ($\sim 10^{15}$ GeV 
\cite{bil}), however much lower neutrino masses ($\sim0(1)$ TeV) can be 
possible in principle  
within the GUT models, too \cite{mink1tev}.

In this Thesis I will investigate two different processes with heavy
neutrinos in the NLC collider for both models mentioned above: 
direct production of heavy neutrinos by the $e^-e^+ \rightarrow
\nu N$ 
process and their indirect detection
by the $e^-e^- \rightarrow W^-W^-$ process. The first process can be realized
in Nature both by Majorana and Dirac neutrinos. I will focus here
on heavy Majorana neutrinos as a more natural case \cite{dir}. 
The second process can appear
only if a neutrino is of the Majorana type (lepton number non-conservation).
Other processes with heavy neutrinos in NLC colliders do not seem  so
promising \cite{belang} with the exception of the $e^- \gamma \rightarrow N W^-$ direct
production process \cite{raid}.

This Thesis is organized as follows. Section 2 
discusses a possible coupling between light and heavy neutrinos. This
parameter is important 
for processes considered in this Thesis. I also define there two distinctive 
models:
`see-saw' and `non-decoupling' ones for which the magnitude of this parameter 
can be
markedly different. Section 3 discusses the 
production of heavy neutrinos in an $e^-e^+$ collision.
Possible CP violation effects
on the cross section are discussed (`see-saw' model) and the most optimistic
values of cross section for heavy neutrino production are derived in the
framework of the `non-decoupling' model. As the magnitude of this cross section
gives hope for a heavy neutrino detection the angular
distribution  of an electron from a heavy neutrino decay is analysed as a 
possible heavy neutrino signal in a detector. Section 4 describes the 
$e^-e^- \rightarrow W^-W^-$ process in which again the CP effects in the 
`see-saw'
model and maximal possible values of cross section (`non-decoupling' model)
are analysed. 
Section 5 presents the conclusions of this Thesis.

Three Appendices show 
how neutrino masses appear both in the RHS and the LR models,
describe details about couplings of neutrinos with particles which 
take part in considered processes and give Feynman rules for
Majorana neutrino interactions, used through the Thesis.
\newpage

\setcounter{equation}{0}
\renewcommand{\theequation}{2.\arabic{equation}}
\section{Light-heavy neutrino mixing angle}
\baselineskip 7 mm

In Appendix B I include relevant couplings of neutrino 
with standard particles which take part in considered $e^-e^+ \rightarrow
\nu N$ and $e^-e^- \rightarrow W^-W^-$ processes.
The mixing matrix elements $K_{Nl}$ and $\Omega_{N\nu}$ (Eqs.(B.5),(B.28))
which are lepton analog of Cabibbo-Kobayashi-Maskawa matrices in the quark
sector are crucial in determining the magnitude of the cross sections. Among
them $K_{Ne}$, coupling of heavy neutrino with electron (positron) is the
most important 
so we have to limit its value as closely as possible. 

Generally two approaches can be applied to fix the $K_{Ne}$ parameter.
The first is historically known as the `see-saw' model \cite{seesaw} and was 
established to explain the small masses of known neutrinos. The second  (let's
call it the `non-decoupling' model) invokes a symmetry argument to explain the 
same
problem. It appears, however, that apart from neutrino mass
explanation other physical variables (as $K_{Ne}$) can differ significantly
in these models.
To show  generally that it is possible, let's discuss the `toy model', in
which  only `light' $(\nu)$
and `heavy' $(N)$ neutrinos exist.
Let us assume that in the $\left( \nu, N \right)^T$ basis 
the neutrino mass matrix is
\begin{equation}
M=\left( \matrix{ a & b \cr b& c } \right).
\end{equation}
For simplicity we assume that  elements a,b,c are real numbers. 
The masses and mixing angle are given by
\begin{equation}
m_{1,2}=\frac{1}{2}\left(a+c\mp\sqrt{(a-c)^2+4b^2} \right),
\end{equation}
and
\begin{equation}
\sin{2\xi}=\frac{2b}{\sqrt{(a-c)^2+4b^2}}.
\end{equation}
There are now two ways to predict the light-heavy spectrum of neutrino masses. 
First is the `see-saw' mechanism where a=0 and $c>>b$ then
\begin{equation}
\mid m_1 \mid \simeq \frac{b^2}{c},\;\;\; \mid m_2 \mid \simeq c>>m_1,
\end{equation}
and unavoidably (Eq.(2.3))
\begin{equation}
\xi \simeq \frac{b}{c} <<1.
\end{equation}
or taking into account relations (2.4)
\begin{equation}
\xi \simeq \sqrt{\frac{\mid m_1 \mid}{m_2}} <<1.
\end{equation}
We can generalize the mass matrix (2.1) to the form as given in the RHS or
the LR models (see Appendix A) and then we can expect that
the light-heavy neutrino mixing $K_{Ne}$ equals 
\begin{eqnarray}
K_{Ne} & \sim & \frac{<m_D>}{<M_R>} \\
\mbox{\rm or} && \nonumber \\ 
K_{Ne} & \sim & \sqrt{ \frac{m_{\nu_e}}{M_N} } .
\end{eqnarray}
Let's stick for a moment to an example in which $m_D$
and $M_R$ matrices are taken in the following form
\begin{eqnarray}
m_D=\left( \matrix{ 1. & 1. & 0.9 \cr
                1. & 1. & 0.9 \cr
                0.9 & 0.9 & 0.95 } \right)\;\;,\;\;
M_R=\left( \matrix{ 10^2 & 0 & 0 \cr
                0 & 10^3 & 0 \cr
                0 & 0 & 5000 } \right).
\end{eqnarray}
\newpage
The neutrino mass matrix $M_{diag}$ which follows (Appendix A) gives
a realistic spectrum of neutrino masses 
($m_{\nu_e} 
=0$ eV, $m_{\nu_{\mu}} \simeq 16\;\mu$eV, $m_{\nu_{\tau}} \simeq 31$ MeV\footnote{
This value is too big when we take into account up-to-date data (Eq.(1.1)).
However, it makes it possible mention one of neutrinos' experiments
which gives an interesting signal lately. An excess of events observed in 
the KARMEN 
detector \cite{karmen} is tentatively interpreted as the decay of 33.9 MeV 
neutral particle. The identification with $\nu_{\tau}$ (SM's, isodoublet)
neutrino is rejected by ALEPH limit of 24 MeV \cite{pdglight}. It should be 
noted,
however, that this limit was questioned lately in the context of neutrino mixing
scenarios \cite{guzzo}. This value can be easily put down to the ALEPH limit 
by some play with parameters in matrices $m_D,M_R$.}) -
light neutrinos\footnote{We can see that this mass
spectrum for light neutrinos is not consistent with cosmological constraints 
where bound $\sum_{light}m_{\nu_{light}}\leq 23$ eV exists \cite{cosm}.
Masses of the light neutrinos can be reconciled with this agreement
in many ways. For instance, we can take $m_D \rightarrow 10^{-3} m_D$
(then interesting mixing angles $K_{Ne}$ will be, however, smaller - in agreement
with Eq.(2.7)) or take 
{\scriptsize $m_D=\left( \matrix{ 1. & 1. & 1. \cr
                1. & 1. & 1. \cr
                1. & 1. & 1.-10^{-6} } \right)\;\;,\;\;
M_R=\left( \matrix{ M_1 & 10^{-6} & 10^{-6} \cr
                10^{-6} & M_2 & 10^{-6} \cr
                10^{-6} & 10^{-6} & M_3 } \right)
$} with relation $\frac{1}{M_1}+\frac{1}{M_2}+\frac{1}{M_3} \simeq0$
\cite{ing}
and $M_{1,2,3} \geq 100$ GeV. Then the mixing angles $K_{Ne}$ will be the same
as for the matrices (2.9).}
and ($M_1=$100 GeV, $M_2=10^3$ GeV,
5000 GeV) - heavy neutrinos. 

Mixing matrix U which diagonalize neutrino mass matrix $M^{\nu}$
(Eqs.(A.11), \newline (A.12)) equals
\newline
$$U=$$
\begin{eqnarray*}
\left( \matrix{ .707 & -.38 i & -.596 i & -.010 & -.001 & .2\cdot10^{-3} \cr
               -.707 & -.38 i & -.596 i & -.010 & -.001 & .2\cdot10^{-3} \cr
               .3\cdot10^{-16} & .84 i & -.537  i &-.010 & -.001 &
.18\cdot10^{-3} \cr
              -.2\cdot10^{-16} & .3\cdot10^{-14} i& .017 i &-1.0 & 
-.3\cdot10^{-5} & .1\cdot10^{-6} \cr
            -.3\cdot10^{-16} & .6\cdot 10^{-15}  i& .002 i&.3\cdot10^{-4} &
-1.0 & 
                                                 .1\cdot10^{-6} \cr
             -.5\cdot10^{-32} & .9\cdot 10^{-16} i & .3\cdot 10^{-3} i&
               .6\cdot 10^{-5}  &             .7\cdot10^{-6} & 1.0 }
                                     \right).
\end{eqnarray*} 
\newpage                                                                       
From this matrix, other submatrices $K$ and $K_R$ (see Appendix B) can be
directly obtained
\begin{eqnarray*}
K&=& 
\begin{array}{c}
\\
  \begin{array}{r}
      \mbox{\rm  \footnotesize light} \\
       \mbox{\rm \footnotesize neutrinos}
  \end{array} 
\rightarrow 
\\
\\
 \begin{array}{r}
      \mbox{\rm \footnotesize heavy}   \\
      \mbox{\rm \footnotesize neutrinos}
 \end{array}
\rightarrow 
\end{array}
{\small \left(
\begin{tabular}{ c|c|c } 
 e & $\mu$ & $\tau$  \\
\hline
        .707 & $-$.707 & $.3\cdot10^{-16}$  \\
              $-$.38 i & $-$.38 i & .84 i  \\
        $-$.596 i & $-$.596 i & $-$.537  i  \\ \hline
       $-$.01 & $-$.01& $-$.01 \\
            $-$.001 & $-$.001 & $-$.001 \\
      $.2\cdot10^{-3}$ & $.2\cdot 10^{-3}$ & $.18\cdot 10^{-3}$         
\end{tabular} \right) }
 \sim 
\begin{array}{l}
\;\;\; \mbox{\rm \scriptsize leptons} \\
\left(
\begin{tabular}{ l } 
 0(1) 
 \\
 \\ \hline
 \\
$\frac{1}{M_N}$
\end{tabular} \right),
\end{array}
\end{eqnarray*}
\begin{equation}
\end{equation}                                                                      

\begin{eqnarray*}
K_R&=& 
\begin{array}{c}
\\
  \begin{array}{r}
      \mbox{\rm  \footnotesize light} \\
       \mbox{\rm \footnotesize neutrinos}
  \end{array} 
\rightarrow 
\\
\\
 \begin{array}{r}
      \mbox{\rm \footnotesize heavy}   \\
      \mbox{\rm \footnotesize neutrinos}
 \end{array}
\rightarrow 
\end{array}
{\small \left(
\begin{tabular}{ c|c|c } 
 e & $\mu$ & $\tau$  \\
\hline
 $-.2 \cdot 10^{-16}$ & $-.3 \cdot 10^{-16}$ & $-.5\cdot10^{-32}$ \\
              $.3\cdot 10^{-14}$ i & $.6 \cdot 10^{-15}$ i & 
             $.9\cdot 10^{-16}$ i  \\
               .017 i & .002 i & $.3 \cdot 10^{-3}$  i \\ \hline
             $-1.0$ & $.3 \cdot 10^{-4}$ & $.6 \cdot  10^{-5}$ \\
            $-.3 \cdot 10^{-5}$ & $-1.0$ & $.7 \cdot 10^{-6}$ \\
             $.1\cdot10^{-6}$ & $.1\cdot 10^{-6}$  & 1.0  
\end{tabular} \right) }
 \sim 
\begin{array}{l}
\;\;\; \mbox{\rm \scriptsize leptons} \\
\left(
\begin{tabular}{ l } 
 0 
 \\
 \\ \hline
 \\
$ 0(1)$
\end{tabular} \right). 
\end{array}
\end{eqnarray*}                                                                       
\begin{equation}
\end{equation}
 
We can see explicitly that we have
\begin{equation}
\left| K_{Ne} \right| \sim \frac{1}{M_N},
\end{equation}
in agreement with our previous estimations (Eq.(2.7)). 

 Let's
note that the values of the $m_D$ matrix elements are justifiable for the LR model
(Eqs.(A.25),(A.26)) as 
\begin{equation}
<m_D> \sim <m^l> \sim 0(1) GeV
\end{equation}
as long as $h \sim \tilde{h}$.
In the RHS model  we can choose the same form of the mass matrices $m_D$ and
$M_R$ as in Eq.(2.9), so the mixing angle $K_{Ne}$ will be the same. 
However, in this model we don't
have so strong theoretical motivation for Eq.(2.13) to hold as in the 
LR model;
that is why some people prefer to estimate light-heavy neutrino mixing 
$(K_{Ne})$ taking into account relation (2.8) which is based on physical quantities.

Anyhow we must be careful because we can get the wrong relation for $K_{Ne}$ parameter 
taking into account Eq.(2.8) as our example shows (in this case $K_{Ne}\sim
\sqrt{\frac{m_{\nu_e}}{M_N}}$ equals zero exactly, in disagreement with its
true value (Eq.(2.10)).

So we have shown that we can find the neutrino mass spectrum in agreement with
terrestrial (or cosmological) constraints and to have the mixing angle between
light and heavy neutrino in the
form given by Eq.(2.12) both for the LR and the RHS models. 
This is what I call classical `see-saw' model and I will use this mixing
angle in numerical calculations.

Secondly, let us assume that $a \neq 0$ and due to internal symmetry 
$ac=b^2$, we have
\begin{eqnarray}
m_1&=&0, \nonumber \\
m_2&=&a+c, 
\end{eqnarray}
and
\begin{equation}
\sin{\xi}=\frac{2\sqrt{ac}}{a+c}.
\end{equation}
If the symmetry which at the tree level gives the relation $ac=b^2$ is broken 
we obtain
$$m_1 \neq 0 << m_2$$ in the higher order (see e.g. \cite{moh}). In this sort of models 
$\sin{2\xi}$ is not connected with the ratio $m_1/m_2$ and can be large 
$( \sin{2\xi} \simeq 1)$ for $a \simeq c$  \cite{9_2},\cite{dir},\cite{witt}.

So for the `non-decoupling' model
the mixing angles are independent parameters not connected 
to the neutrino masses and are only bound by existing experimental data. 
Constraints come from:

(i) Low energy experiments (e.g. lepton universality, the $\mu$ decay)
and the LEPI give also information about heavy neutrinos with masses above $M_Z$.
The reason is that due to unitarity properties of the U matrix (Eq.(A.11)), 
the nonzero mixing 
matrix elements $K_{Ne}$ (Eqs.(B.5),(B.28)) slightly reduce the couplings 
of light neutrinos from their SM 
values thus affecting all processes including these particles \cite{shr} 
(in the SM matrix U can be taken as I and then matrix K in Eq.~(B.3)) equals
I, too). The up-to-date
limit for the RHS model is \cite{9_6} 
\begin{equation}
\sum\limits_{N(heavy)}K_{Ne}^2 \leq \kappa^2=0.0054.
\end{equation}
Practically the same limit exists for the LR model (\cite{nardi}).

(ii) The lack of signal of neutrinoless double-$\beta$ decay $(\beta\beta)_{0\nu}$ gives
the
bound for light neutrinos
\begin{equation}
\left| \sum_{\nu(light)}K_{\nu e}^2m_{\nu} \right| <
\kappa^2_{light}
\end{equation}
where $\kappa^2_{light}<0.65$ eV \cite{klight}. \\

(iii) From the $\left( \beta\beta \right)_{0\nu}$ process it is also possible
to get the bound for heavy neutrinos\footnote{This bound can be obtained 
when more than one nuclei is considered. Then because nuclear matrix elements
differ from nucleus to nucleus, possible cancelations in amplitude between
light and heavy neutrinos' contributions can not appear in all nuclei 
simultaneously \cite{verg}.} 

\begin{equation}
\left| \sum\limits_{N(heavy)}K_{Ne}^2\frac{1}{M_N} \right| < \omega^2.
\end{equation}
The up-to-date value is $\omega^2<(2-2.8)\cdot10^{-5}\rm\;TeV^{-1}$
for the RHS model \cite{omega} and similarly for the LR model \cite{hirlr}
\footnote{Some people claim \cite{hmraid} that this value
is too strict and can be loosen by a factor $\sim 40$. It does not change, 
however,
any other conclusions given there quantitatively. The most important for our 
discussion is
that such bound can be derived from experimental data (the situation when
the relation (2.18) is not taken into account were considered in
\cite{pr6}.}. \\

The last constraint which I use to fix $K_{Ne}$ comes from the fact that the mass term for
the left-handed neutrinos is absent. Then (Eqs.(A.9), \newline
(A.11),(A.12))

(iv) 
\begin{equation}
\sum\limits_{\nu (light)}K_{\nu e}^2m_{\nu}+\sum\limits_{N(heavy)}K_{Ne}^2M_N 
=M_L \equiv 0.
\end{equation}

This fact confronted with Eq.(2.17) gives
\begin{equation}
\left| \sum_{N(heavy)}K_{Ne}^2M_N \right| < \kappa^2_{light}.
\end{equation}

This relation includes an interesting information. To get meaningful
values of cross sections for the studied processes as large as possible 
values of $K_{Ne}$ are needed. As $\kappa^2_{light}$ is very small
the only possibility to reconcile these two facts is to assume that some
$K_{Ne}$ matrix elements are complex numbers. If CP symmetry is conserved
then neutrinos with purely complex $K_{Ne}$ numbers have opposite $\eta_{CP}$
parities to neutrinos with real ones \cite{bil}. Relation
(2.19) was important to get this conclusion. 
However, we know that Majorana neutrinos get mass through
radiative corrections \cite{moh}, \cite{radneut} 
and consequently Eq.(2.19) must be modified

\begin{equation}
\sum\limits_{\nu (light)}K_{\nu e}^2m_{\nu}+\sum\limits_{N(heavy)}K_{Ne}^2M_N 
=M_L^r
\end{equation}

We don't know in principle how large $M_L^r$ can be. In what follows I will
hold this general formula with $M_L^r$.

Taking into account relations (2.16)-(2.21) the largest possible $K_{Ne}$ 
depends on the
number of right-handed neutrinos $n_R$ ($n_R=3$ in the LR model)

$\bullet\;n_R=1$ or all heavy neutrinos with the same CP parities \\

In this case very restrictive constraint results from Eqs.(2.17),(2.21)
\begin{equation}
\left| M_L^r- \sum\limits_{N}K_{Ne}^2M_N \right| < \kappa_{light}^2.
\end{equation}
For $M_L^r \simeq 0$ mixing angle $K_{Ne}$ is extremely small.
Another  strict constraint follows from Eq.(2.18) 
\begin{equation}
\mid K_{Ne}^2 \mid < \omega^2 M.
\end{equation}

$\bullet\;n_R=2$ \\

In agreement with the previous discussion we assume that both
heavy neutrinos have opposite CP parities. Let us take
$\eta_{CP}(N_1)=-\eta_{CP}(N_2)=i$. If we denote
$K_{N_1e}=x_1,\;K_{N_2e}=ix_2,\;m_1=M,\;m_2=AM$ $(A>1)$ then from
Eqs.(2.16)-(2.18),(2.21) we get
\begin{eqnarray}
x_1^2+x_2^2 & \leq & \kappa^2 \\
\left| \frac{M_L^r}{M}-x_1^2 +Ax_2^2 \right| & \leq & \frac{\kappa^2_{light}}{M} \\
\left| x_1^2- \frac{x_2^2}{A} \right| & \leq & \omega^2 M.
\end{eqnarray}

In Eq.(2.25)  $\delta \equiv \frac{\kappa^2_{light}}{M}$ is very small for $M>100$ GeV
$(\delta<10^{-11})$. Since we want to have values of 
$x_1^2$ and/or $x_2^2$ mixing angles as large as possible then 
$\delta << x_1^2,Ax_2^2$, and $\delta$ parameter
can be practically neglected. It gives $(\frac{M_L^r}{M} \equiv \Delta^r)$
\begin{equation}
x_1^2=\Delta^r+Ax_2^2
\end{equation}
and from inequalities (2.24) and (2.26) we get that masses and mixing angles of heavy neutrinos
with electron must satisfy the following relations
\begin{equation}
x_1^2 \leq A\frac{\kappa^2-\Delta^r}{A+1}+\Delta^r \;\;\;{\rm
and}\;\;\;
x_1^2 \leq \frac{A^2\omega^2 M-\Delta^r}{A^2-1}
\end{equation}
and
\begin{equation}
x_2^2 \leq \frac{\kappa^2-\Delta^r}{A+1}\;\;\;{\rm and}\;\;\;
x_2^2 \leq
\frac{A}{A^2-1}(\omega^2 M-\Delta^r ).
\end{equation}
As for masses
$0.1\; {\rm TeV} < M < 1 {\rm TeV}, \; \kappa^2 \gg \omega^2
M$, the
second inequalities are usually stronger. The only possible way to
get large $x_1^2$ is to assume that $A \rightarrow 1$. 
We can derive some information about $M_L^r$, too. From Eqs.(2.28),(2.29)
we have $\Delta^r \leq \omega^2 M$, so for given value 
$\omega^2$  we have
\begin{equation}
M_L^r \leq 10\; \mbox{\rm MeV}.
\end{equation}
However, for $M \geq 1$ TeV restrictions weaken very fast.
The largest possible values of the mixing matrix elements are 
$(A \rightarrow 1)$
\begin{eqnarray}
K_{N_1e}^2 & \rightarrow & \frac{\kappa^2}{2},  \nonumber \\
K_{N_2e}^2 & \rightarrow & \frac{\kappa^2}{2}.
\end{eqnarray}
Neutrinos with those mixing angles have opposite CP parities, so Dirac neutrino is realized there.

$\bullet\; n_R=3$ \\

Let's assume that $\eta_{CP}(N_1)=\eta_{CP}(N_2)=-\eta_{CP}(N_3)=i$.
If we denote $K_{N_1e}=x_1,\; K_{N_2e}=x_2,\; K_{N_3e}=ix_3$ and
$m_1=M,\; m_2=AM,\; m_3=BM$, then relations (2.16)--(2.21) give a set of
inequalities. I consider the more interesting case $A>B$ (for $A<B$
the mixing parameters are much smaller)
in which the following inequalities are satisfied
\begin{equation}
x_2^2 \leq -x_1^2\frac{1+B}{A+B}+\left(\kappa^2+\frac{\Delta^r}{B}
\right) \frac{B}{A+B},
\end{equation}
\begin{equation}
x_2^2 \geq x_1^2\frac{B^2-1}{A^2-B^2}A-\left(
\omega^2M-\frac{\Delta^r}{B^2} \right) \frac{AB^2}{A^2-B^2},
\end{equation}
and
\begin{equation}
x_2^2 \leq x_1^2 \frac{B^2-1}{A^2-B^2}A+ \left( \omega^2
M+\frac{\Delta^r}{B^2} \right) \frac{AB^2}{A^2-B^2}.
\end{equation}
$x_3^2$ can be found from the relation
\begin{equation}
x_3^2=\frac{1}{B} \left( x_1^2+Ax_2^2-\Delta^r \right) .
\end{equation}

From inequalities (2.32)-(2.34) we can find a region in the
$(x_1^2,x_2^2)$ plane of still acceptable mixing parameters. The
region (which is schematically shown in Fig.~1) depends on the
chosen values of $M,A$ and $B$.

Maximal values of mixing matrix elements are ($A \rightarrow \infty $ )
\begin{eqnarray}
x_1^2 & \rightarrow & \frac{\kappa^2+BM\omega^2}{1+B} \nonumber \\
x_2^2 & \rightarrow & 0 \nonumber \\
x_3^2 & \rightarrow & \frac{B \kappa^2-BM\omega^2}{1+B} .
\end{eqnarray}

We can see that in this case a heavier neutrino can have a larger mixing
angle than the lightest one.

Let's stress once more that quantitatively  
restrictions given in Eqs.(2.32)-(2.35) are true both for the LR and the RHS 
models as $\kappa^2$, 
and $\omega^2$ are practically the same in both models.

\newpage

\ \\

\vspace{6.5 cm}
\begin{figure}[h]
\includegraphics{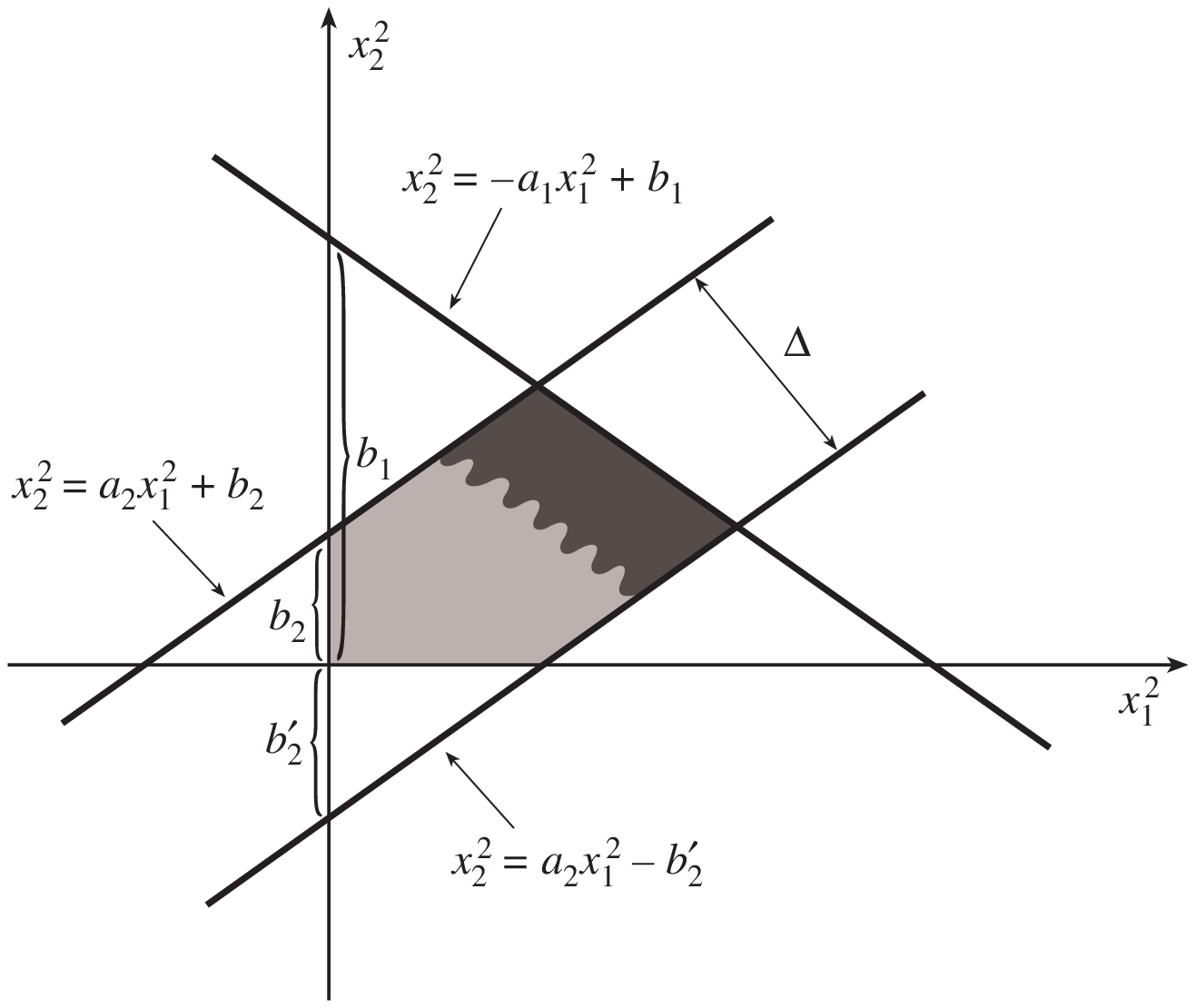}
\end{figure}

\baselineskip 5 mm
{\footnotesize Fig.1. Sketch of the region in $x_1^2-x_2^2$ plane of still experimentally
acceptable mixing
parameters. I use the following denotations (see Eqs.~(2.32)-(2.34) in the text) \newline
$\begin{array}{lll}
a_1=\frac{1+B}{A+B},& b_1=\left( \kappa^2+\frac{\Delta^r}{B} \right)
\frac{B}{A+B},& a_2=A\frac{B^2-1}{A^2-B^2} \\
b_2=\left( \omega^2M-\frac{\Delta^r}{B^2} \right) \frac{AB^2}{A^2-B^2},&
b_2'=\left( \omega^2M+\frac{\Delta^r}{B^2} \right) \frac{AB^2}{A^2-B^2}&
\end{array}$ \newline
For masses $M<1$ TeV, $b_2 \sim b_2' \ll 1$ and the region is very narrow
$\left( \Delta \rightarrow 0 \right)$. The more shadowed region is the place
where the mixing angles are the largest.}
\\

\baselineskip 7 mm

\newpage

\setcounter{equation}{0}
\renewcommand{\theequation}{3.\arabic{equation}}
\baselineskip 7 mm

\section{Direct heavy neutrinos production in $e^-e^+$ collision:
$e^-e^+ \rightarrow \nu N$ process}
\subsection{Which Feynman diagrams are the most important?}

At the tree level the $e^-e^+ \rightarrow N_aN_b$ process proceed through the following
Feynman diagrams (Fig.2). 
\vspace{6.5 cm}
\begin{figure}[h]
\includegraphics{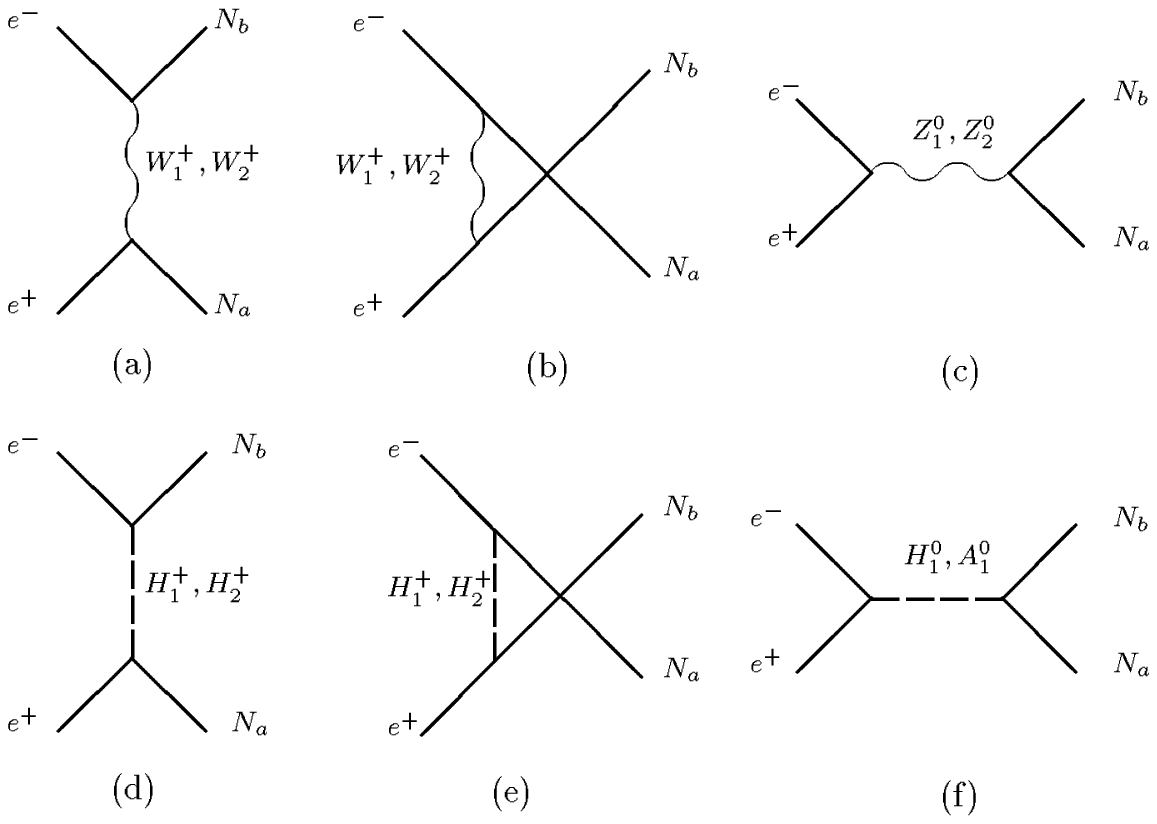}
\end{figure}

{\footnotesize Fig.2 $e^-e^+ \rightarrow N_a N_b$ process at the
tree level in the framework of the LR model.} \\

$N_a$ and $N_b$ stand for any neutrinos - light or heavy.
Denotations are given for the LR model
where we have two pairs of charged and neutral gauge bosons and couple
of Higgs particles (only Higgs particles with nonzero couplings to
electrons are depicted there - see Appendix~B for details). In the framework
of the RHS model
only two first diagrams contribute (with only one charged gauge
boson $W^-_1 \equiv  W^-$). 
I will consider Majorana neutrinos, so crossing diagrams are taking into
account.

The t- and u-channels are the most important for this process 
with the $W_1$ exchange. Other, like $Z_{1(2)}$ contributions, are
negligible because of off-peak energies ($\sqrt{s}>>M_{Z_1}$) and a large mass
of additional gauge boson $Z_2$ ($\sqrt{s}<<M_{Z_2}$) - see Eq.(B.16). 
The $W_2$ gauge boson couples with an electron predominantly by the right-handed
currents (see Eqs.(B.18) and (B.21)). 
Large mass of $W_2$ decide, however, about smallness
of the $W_2$ contribution to the process. To be specific, 
results for $\sqrt{s}=500$ GeV collider are summarized in Table 1 where
the case of two heavy neutrinos production is included, too (with
contribution dominated by the $W_2$ exchange).
\begin{table}
\centering
\begin{tabular}{c c c}
\cline{1-3}
\ &\ &\ \\
$e^+e^-$ &\ $ \rightarrow \nu N$ & \ $ \rightarrow NN$  \\
\cline{1-3}
$W_1$ &  {\bf $\sim 0.2\%$} & $\sim 99.5\%$ \\
\ &\ &\ \\
$Z_1$ & $\sim 97.5\%$ & $\sim 99.5\%$ \\
\ &\ &\ \\
$W_2$ & $\sim 99.9\%$ & {\bf $\sim 2\%$} \\
\ &\ &\ \\
$Z_2$ & $\sim 99.9\%$ & $\sim 99.2\%$ \\
\ &\ &\ \\
\cline{1-3}
\end{tabular}
\caption{{\footnotesize The contributions of different diagrams with $W_{1,2}$ and $Z_{1,2}$
exchange to the total cross section. The numbers in the table present the part of
${\sigma}_{total}$ after removing corresponding diagrams.}}
\end{table}
That is why results for the LR and the RHS models
are practically the same for the single heavy neutrino production in
the $e^-e^+$ collision. 

Apart from
additional gauge bosons in the LR model additional contributions to this
process come from Higgs particles' exchange (Fig.2). 
I use the unitary gauge so diagrams with Goldstone particles
exchange are not taken into account. How much can they change
results? The precise values of all couplings are presented in Appendix B.
In the approximation used here $\left( v_R \gg y\;,\;m_e\simeq0 \right)$
(Eq.(B.29)) only two neutral Higgses couple in the s channel. 
The lightest Higgs particle $H_0^0$ (equivalent of the SM's one) couples 
to the
$e^-e^+$ proportionally to the electron mass and its effect is negligible in the 
energy range which I consider (see Eq.(B.35)). 
The influence of two charged $H_{1,2}^+$ and two
neutral $\left( H_1^0,A_1^0 \right)$ Higgs particles is not obvious. At first
sight
their coupling, even to the light leptons $\left( e^-e^+,e\nu\right), $ can be large
as there are terms in the vertex proportional to heavy neutrinos mass. They are, 
however, multiplied by the mixing matrices which can have small terms so the 
total effect needs precise analysis.

For $M_{W_1}=80$ GeV and $M_{W_2}=1600$ GeV
\cite{mw21600} we can find (Eqs.(B.15-16)) $y\simeq250$ GeV, $v_R \simeq 3500$ GeV and then
for $\epsilon =0$ all Higgs boson masses are of 2.5 TeV order (Eq.(B.30))
\begin{equation}
M_{H_1^{\pm}}\ \simeq M_{H_2^{\pm}} \simeq M_{H_1^0} \simeq M_{A_1^0} \simeq
\;2450\;\mbox{\rm GeV}.
\end{equation}
For the neutral Higgs bosons $H_1^0$ and $A_1^0$ the  masses 
of order 2.5 TeV are not large enough to reduce the $\bar{K}^0-K^0$ transition.
To generate proper mass splitting in $\bar{K}^0-K^0$ system it was found that 
masses of neutral Higgs particles must be above 10 TeV.
Otherwise these particles could have caused that flavour changing neutral currents
(FCNC) were too large \cite{fcnc1},\cite{fcnc2}.

There are two ways of obtaining such large masses of neutral Higgs bosons. 
Firstly we can assume that some parameters which are present in Higgs potential
are very large (greater than 1). Then minimalization of the Higgs potential can
give larger masses of Higgs particles \cite{pr3}. 

Secondly, we can avoid the fine tuning problem for the Higgs parameters 
mentioned above by taking 
the same VEV $\kappa_1\simeq \kappa_2$, so then $\epsilon \simeq 1$ (Eq.(B.30)) 
and the masses 
$M_{H_1^0}, M_{A_1^0} $ and $M_{H_2^{\pm}}$ are much greater than $\frac{1}{2}
v_R^2$, so greater than 10 TeV.

In the first case the presence of large masses in the propagator causes that the total
contribution of the $H_1^0$ and $A_1^0$ exchange in the s channel is very small.
When $\epsilon \rightarrow 1$, the couplings of neutral Higgses 
$H_1^0$ and $A_1^0$ to leptons become stronger (see Eqs.(B.34)-(B.36)) and compensate the 
influence of the propagator. The total effect depends on the additional vertex 
contributions given in Eqs.(B.35) and (B.36). I have calculated numerically the factors 
of the type
\begin{equation}
(K^{\dagger}M^{\nu}_{diag}K_R)_{ee}\;\;,\;\;(Km_l^{diag}K_R^{\dagger})_{ab}\;\;,\;\;
(\Omega M^{\nu}_{diag})_{ab}
\end{equation}
which are present in those couplings for different values of the 
heavy neutrino masses. $K,K_R,\Omega$ and $M_{diag}^{\nu}$ are taken as 
discussed in Section 2 (Eqs.(2.9)- \newline 
(2.11),(B.28)). Then the factors 
are of the same order independently of the neutrino masses.
It is caused by the fact that for larger neutrino masses the appropriate
mixing matrix elements are smaller. I have checked that the influence of the 
scalar $H_1^0$ and pseudoscalar $A_1^0$ exchange diagrams on the total cross 
section is completely
negligible for considered energy range. I have checked also the contribution of the charged Higgs bosons exchange
in the t-u channels. 
In Table 2 the ratios of cross sections with only gauge bosons
($\sigma_{gauge}$) or Higgs particles ($\sigma_{Higgs}$) to $\sigma_{total}$
in which all Feynman diagrams are taken into account are presented. 
We can see that Higgses have
no meaning for heavy Majorana neutrino production. For $\nu_{\mu}N$ and
$\nu_{\tau}N$ neutrino production the Higgs exchange mechanism gives only the 
contribution of the order of $10^{-4}$.
\begin{table}
\centering
\begin{tabular}{c c c}
\cline{1-3}
  &\ &\ \\
$e^-e^+$ &\  $\sigma_{gauge}/\sigma_{total}$ &\ $\sigma_{Higgs}/\sigma_
{total}$ \\
 &\ &\ \\
\cline{1-3}
$\rightarrow \nu_eN(100)$ & $\simeq 100\%$ & $\simeq .0001\%$ \\
  &\ &\ \\
$\rightarrow \nu_{\mu}N(100)$ & $\simeq 100\%$ & $\simeq .01\%$ \\
  &\ &\ \\
$\rightarrow \nu_{\tau}N(100)$ & $\simeq 100\%$ & $\simeq .01\%$ \\
\cline{1-3}
\end{tabular}
\caption{{\footnotesize The contribution of the gauge and Higgs bosons to the total cross section
for LEPII energy.}}
\end{table}
Moreover, these results are not sensitive to the $\epsilon$ factor. The Feynman diagram 
with $H_1^+$ exchange (which is the most important)  is not sensitive 
to this factor at all (Eqs.(B.30),(B.31)).
 The $H_2^+$ exchange diagram is sensitive to this factor and the
contribution to the total cross section increases with increasing $\epsilon$
(Eqs.(B.32),(B.33)),
but as the propagator for this particle is sensitive to this factor, too
(Eq.(B.30)),
the increase is rather small and even for $\epsilon=1$ does not predominate 
the $H_1^+$ contribution. The result is shown in Fig.3 for the $e^-e^+
\rightarrow \nu_{\tau}N$ ($M_N=100$ GeV) process.

\vspace{6.5 cm}
\begin{figure}[h]
\includegraphics{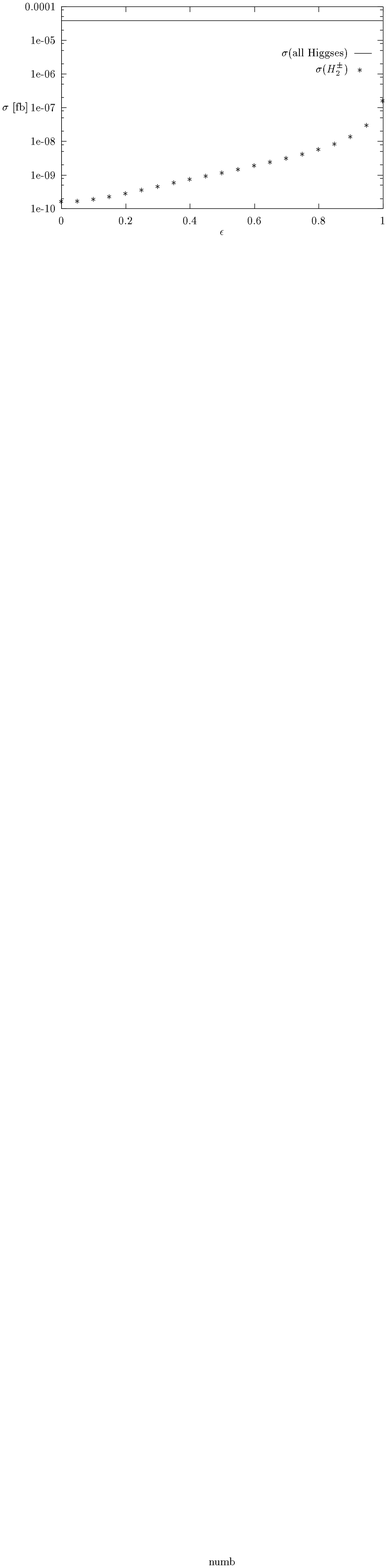}
\end{figure}

\baselineskip 5 mm

{\footnotesize Fig.3.~The $\epsilon$ dependence for $\sigma [e^-e^+
\rightarrow \nu_{\tau} N(100) ]$ cross section given by all physical Higgs
boson exchange (solid line) and only by $H_2^+$ Higgs boson exchange
(asterisk line) for LEPII energy.}

\baselineskip 7 mm


%
%

To sum up, two first diagrams from Fig.2 give the largest contribution to
the single heavy neutrino production process ($W_1$ exchange) and to the pair
of heavy neutrino production process ($W_2$ exchange). Helicity amplitudes
for these diagrams are gathered in Eqs.~(3.3)-(3.5) (full details with all Feynman
diagrams Fig.2~(a)-(f) are given in \cite{pr3}) where 
$\sigma, \bar{\sigma}, \lambda_{1,2}$ denote helicities of electron, positron
and two heavy neutrinos respectively, $\beta_{1,2}=\frac{q}{E_{1,2}}$
are kinematical factors with q - momentum, $E_{1,2}$ - energy of produced
neutrinos in CM frame and index i=1,2 describes two charged gauge bosons
 
\begin{equation}
-iM(\sigma, \bar{\sigma};\lambda,\bar{\lambda}) \propto 
\frac{A_1^{t_i}}{t-M_{W_i}^2}-\frac{A_1^{u_i}}{u-M_{W_i}^2},
\end{equation}

\begin{eqnarray}
A_1^{t_i}\left( \sigma,\bar{\sigma};\lambda_1,\lambda_2 \right)&=&
{\left(A_L^{i}\right)}^{\ast}_{1e} {\left(A_L^{i}\right)}_{2e}
{\delta}_{\Delta\sigma,-1} \sqrt{( 1-2\lambda_1\beta_1)
(1+2{\lambda_2}\beta_2) } \ \ \ \ \ \ \ \ \ \ \ \ \ \ \ \nonumber \\
&+&{\left(A_R^{i}\right)}^{\ast}_{1e} {\left(A_R^{i}\right)}_{2e}
{\delta}_{\Delta\sigma,+1} \sqrt{( 1+2\lambda_1\beta_1)
(1-2{\lambda_2}\beta_2) },\\
&& \nonumber \\
A_1^{u_i}\left( \sigma,\bar{\sigma};\lambda_1,\lambda_2 \right)&=&
{A_1^{t_i}}^{\ast}\left(\lambda_1 
\longleftrightarrow {\lambda_2} ,\beta_1 \longleftrightarrow \beta_2 \right)
\nonumber \\
&=&{\left(A_L^{i}\right)}_{1e} {\left(A_L^{i}\right)}_{2e}^{\ast}
{\delta}_{\Delta\sigma,-1} \sqrt{( 1-2\lambda_2\beta_2)
(1+2{\lambda_1}\beta_1) } \ \ \ \ \ \ \ \ \ \ \ \ \ \ \ \nonumber \\
&+&{\left(A_R^{i}\right)}_{1e} {\left(A_R^{i}\right)}_{2e}^{\ast}
{\delta}_{\Delta\sigma,+1} \sqrt{( 1+2\lambda_2\beta_2)
(1-2{\lambda_1}\beta_1) },
\end{eqnarray}
 
Factors $A_{L,R}^i$ are defined in Eq.(B.21).
As we can see these factors in principle are not real and their
complex phases are connected with CP violation effects in the lepton sector.
As their complexity caused destructive or constructive
interferences among different terms in Eqs.(3.3)-(3.5) it is worth to investigate
the influence of CP phases on the considered reaction.

\subsection{CP violation effects in the $e^-e^+ \rightarrow N_1 N_2$ process (`see-saw'
model)}

CP violation have been observed until now only as a small effect 
in the $K^0-\bar{K^0}$ system (quark sector).  Smallness of this effect
is understood in the framework of the SM. Physical quantities must be 
independent of the choice of weak basis so only weak basis invariants
enter in a measurable quantity. It can be shown
that for 3 families there is only one independent invariant which can be built
of quark mass matrices $M_u,M_d$ \cite{weakinv}
\begin{eqnarray}
\{\mbox{\rm weak}\;\mbox{\rm invariant}\}&=&-2i(m_t^2-m_c^2)(m_t^2-m_u^2)(m_c^2-m_u^2)
(m_b^2-m_s^2) \nonumber \\
&&(m_b^2-m_d^2)(m_s^2-m_d^2)Im ( V_{cd}V_{ub}V_{cb}^{\ast}V_{ud}^{\ast})
\end{eqnarray}
where $V_{ij}$ are elements of the Cabibbo-Kobayashi-Maskawa mixing matrix.
Because quark masses are non-degenerate then in the SM all CP violating
effects are proportional to
\begin{equation}
\delta_{KM}= Im ( V_{cd}V_{ub}V_{cb}^{\ast}V_{ud}^{\ast}) .
\end{equation}
Using the unitarity of the CKM matrix we can write
\begin{equation}
\mid \delta_{KM} \mid = \mid Im (V_{cd}V_{ub}V_{cb}^{\ast}V_{ud}^{\ast}) \mid
\end{equation}
and substituting experimental values we have
\begin{equation}
\mid \delta_{KM} \mid < {10}^{-4}.
\end{equation}
This tiny quantity is responsible for CP violation effect in $K^0-\bar{K}^0$ system.

No CP violation has been observed in the lepton sector. This can be
related to the smallness of the masses of the known neutrinos.
Heavy Majorana neutrinos can potentially change this situation.

CP violation effects can be caused by complex phases in the mixing matrices
$K$ and $K_R$.  
However not all phases in the mixing matrices are CP violating. Some of them
can be eliminated by redefinition the fermion fields. For
instance, in the quark sector of the SM, six phases which define $3 \times 3$
unitary mixing matrix reduce to one phase after appropriate fermion fields
redefinition \cite{ckm}. Let's check how does it work for considered the 
LR model.

The relevant 
parts of the model's Lagrangian for studying the CP properties are the charged-current 
interaction and the lepton mass Lagrangian.
They are given by (weak basis)
\begin{equation}
L_{CC}=\frac{g}{\sqrt{2}} \left( \bar{\nu}_L\gamma^{\mu}l_LW_{L\mu}^++\bar{\nu}_R
\gamma^{\mu}l_RW_{R\mu}^+ \right) + h.c.
\end{equation}
and (Eqs.(A.21),(A.24-26))
\begin{equation}
L_{mass}=-\frac{1}{2} \left( \bar{n}_L^cM_{\nu}n_R+\bar{n}_RM_{\nu}^{\ast}
n_L^c \right) - \left( \bar{l}_Lm^ll_R+\bar{l}_R{m^l}^{\dagger}l_L \right)
\end{equation}
where $n_R$ is a six-dimensional vector of the neutrino fields
\begin{eqnarray}
n_R&=&\left( \matrix{ \nu_R^c \cr
                      \nu_R } \right)\;,\; \nu_R^c=i\gamma^2\nu_L^{\ast}, 
     \nonumber \\
n_L&=&\left( \matrix{ \nu_L \cr
                      \nu_L^c } \right)\;,\; \nu_L^c=i\gamma^2\nu_R^{\ast}.
\end{eqnarray}            
The most general CP transformation which leaves the gauge interaction
(3.10) invariant is \cite{redef}
\begin{eqnarray}
l_L &\rightarrow & V_L Cl_L^{\ast}\;,\;\;\nu_L \rightarrow V_L C\nu_L^{\ast},
\nonumber \\
l_R &\rightarrow & V_R Cl_R^{\ast}\;,\;\;\nu_R \rightarrow V_R C\nu_R^{\ast}.
\end{eqnarray}
where $V_{L,R}$ are $3 \times 3$ unitary matrices acting in
lepton flavour space and C is the Dirac charge conjugation matrix. 
For the full Lagrangian to be invariant under (3.13) the lepton mass matrices 
$m_D,M_R$ and $m^l$ have to satisfy the conditions
\begin{eqnarray}
V_L^{\dag}m_DV_R&=&m_D^{\ast}, \nonumber \\
V_R^TM_RV_R&=&M_R^{\ast}, 
\end{eqnarray}
and 
\begin{eqnarray}
V_L^{\dag}m^lV_R&=&{\left( m^l \right) }^{\ast} .
\end{eqnarray}
The relations expressed by Eqs.(3.14) and (3.15) are weak-basis independent and constitute
necessary and sufficient condition for CP invariance. It means that if for
given matrices $m_D,M_R$ and $m^l $, there exist two unitary matrices $V_L$ and
$V_R$ such that relations (3.14),(3.15) hold then the model is CP invariant 
and, on the other hand, if CP is the symmetry of the model then such matrices $V_L$ and 
$V_R$ exist.
The most convenient basis for studying CP symmetry is the weak basis in which
charged lepton mass matrix $m^l$ is real, positive and diagonal \cite{zrag}
\begin{equation}
m^l=diag[m_e,m_{\mu},m_{\tau}].
\end{equation}
Then for non-degenerate, non-vanishing $m_e \ne m_{\mu} \ne m_{\tau}$ 
Eq.(3.14) and (3.15)
imply that matrices $V_{L,R}$ are diagonal and equal 
\begin{equation}
V_L=V_R=diag[e^{i\delta_1},e^{i\delta_2},e^{i\delta_3}].
\end{equation}
From Eqs.(3.14) and (3.15) then it follows that the model has CP symmetry if and only if the
matrices $m_D$ and $M_R$ have the elements
\begin{eqnarray}
(m_D)_{ij}&=&\mid (m_D)_{ij} \mid e^{+\frac{i}{2}(\delta_i-\delta_j)},
\nonumber \\
(M_R)_{ij}&=&\mid (M_R)_{ij} \mid e^{-\frac{i}{2}(\delta_i+\delta_j)}
\end{eqnarray}
in the basis where $m^l$ is diagonal. 
Altogether we have six CP-violating phases 
($\frac
{n(n+1)}{2}=$ for symmetric $M_R$ plus $\frac{n(n-1)}{2}$ for hermitian
$m_D$ give totally $n^2$ phases minus n phases connected with fields
redefinition) which can be written in the form

\begin{eqnarray}
M_R&=&\left( \matrix{ M_{11}e^{i\alpha_1} & M_{12} & M_{13} \cr
                    M_{12} & M_{22}e^{i\alpha_2} & M_{23} \cr
                    M_{13} & M_{23} & M_{33}e^{i\alpha_3} } \right), 
\nonumber \\
m_D&=&\left( \matrix{ m_{11} & m_{12}e^{i\beta_1} & m_{13}e^{i\beta_2} \cr
                    m_{12}e^{i\beta_1} & m_{22} & m_{23}e^{i\beta_3} \cr
                    m_{13}e^{i\beta_2} & m_{23}e^{i\beta_3} & m_{33} } \right).
\nonumber
\end{eqnarray}


For energies much bigger than the masses of neutrinos $N_1$ and $N_2$
($\sqrt{s}>>0(1)$ TeV) or when $\sqrt{s} \sim 0(1)$ TeV but with at least
one light neutrino, the t channel predominantly contributes to the
M$(-+;-+)$
amplitude (left-handed current) and M$(+-;+-)$ one (right-handed current)
and the u channel gives contributions to M$(-+;+-)$ and M$(+-;-+)$ amplitudes 
(see Eqs.(3.4)-(3.5)). We can see that in these cases there is no interference 
between t and u channels.

For the energy just above the production threshold there is no helicity
suppression mechanism for two heavy neutrino production process $e^-e^+
\rightarrow N_1N_2$ (Eqs.(3.4)-(3.5)) and final neutrinos with all helicity
states can be produced by each channel diagram. These are the best
conditions for observing the CP violation effects because in this case t and
u channels contribute to the same helicity states. 
As heavy
neutrinos predominantly couple to $W_2$ gauge bosons (see matrix U (Eq.(2.11)
with
  ${(K_R)}_{{N_4}e}=1$ and Eq.(B.21)), the 
diagrams with $W_2$ (right-handed currents)
  are the most interesting (that is the reason why we don't consider
  results for the RHS case at all as they are small in comparison to the
  LR model ones). 
  
  Another question is in what experimental observables the CP effects 
  are visible. From the discussion presented above we can see that they can be 
  looked for in polarized angular distribution. 
  And what
  about the unpolarized angular distribution? If CP is conserved then the
  helicity amplitude satisfies the relation ($\Theta$ and $\phi$ are CM 
  scattering angles)
  
\begin{eqnarray}
M(\sigma,\bar{\sigma};\lambda_1,\lambda_2;\Theta,\phi)&=&-\eta_{CP}^{\ast}(1)
\eta_{CP}^{\ast}(2)\times \nonumber \\
&&M(-\bar{\sigma},-\sigma;-\lambda_1,-\lambda_2;\pi-\Theta,\pi+\phi).
\end{eqnarray}
If we sum over all
helicity states the unpolarized angular distribution has 
forward-backward isotropy

\begin{equation}
\frac{d\sigma}{d\Omega}(\Theta,\phi)=\frac{d\sigma}{d\Omega}(\pi-\Theta,
\pi+\phi).
\end{equation}
Does it mean that anisotrophy can be observed if CP is violated? Unfortunately not, at least
if we neglect the final state interaction. Without final state interaction from 
CPT symmetry we can prove the relation
\begin{eqnarray}
M(\sigma,\bar{\sigma};\lambda_1,\lambda_2;\Theta,\phi)&=&-\eta_{CP}(1)
\eta_{CP}(2) e^{2i(\sigma-\bar{\sigma})(\pi+\phi)}\nonumber \\
&&M^{\ast}(-\bar{\sigma},-\sigma;-\lambda_1,-\lambda_2;\pi-\Theta,\pi+\phi)
\end{eqnarray}
from which the forward-backward isotrophy also follows \cite{petc}.
So the only observables where we can try to find the CP violation effect are the total
cross sections and the polarized angular distributions. As the magnitudes of
the total cross
sections are larger in comparison with polarization angular
distribution I will investigate now the total cross section 
$e^-e^+ \rightarrow N_1N_2$.

There are six phases which cause the CP symmetry breaking. 
I take the matrices $m_D$ and $M_R$ in the form
\begin{eqnarray}
m_D&=&\left( \matrix{ 1. & 1. & .9 \cr
                      1. & 1. & .9 \cr
                      .9 & .9 & .95 } \right),
\end{eqnarray}
and                       
\begin{eqnarray}
M_R&=&\left( \matrix{ 150e^{i\alpha} & 10 & 20 \cr
                     10 & 200e^{i\beta} & 10 \cr
                     20 & 10 & 10^4e^{i\gamma} } \right),
\end{eqnarray}
which produce a reasonable spectrum of light neutrinos (see Section 2). If we compare these
matrices with Eq.(3.18) we can see that if only one or more phases ($\alpha,\beta$ or 
$\gamma$) are not equal 0 or $\pi$ the CP is violated. Two heavy neutrinos with 
masses $M_1\simeq 150 $ GeV and $M_2 \simeq 200$ GeV, almost independent
of the phases $\alpha,\beta$ and $\gamma$, result from this mass matrix.
The appropriate mixing matrix elements $(K,K_{R})_{1e},(K,K_{R})_{2e}$
depend on the phases $\alpha$ and $\beta$ and are
almost independent of the phase $\gamma$. For $\alpha=\beta=\gamma=0$
two neutrinos have equal CP parity and CP is conserved for
\begin{equation}
\eta_{CP}(N_1)=\eta_{CP}(N_2)=+i.
\end{equation}                   
For $\alpha=\pi,\beta=\gamma=0$ CP is also conserved if we introduce the CP
parities  
\begin{equation}
-\eta_{CP}(N_1)=\eta_{CP}(N_2)=+i.
\end{equation}                   
For any other values of phases CP is violated. The production cross sections
versus energy are presented in Fig.4.

Two factors affect the behaviour of the cross section. 
Different $\alpha,\beta,\gamma$ in the matrix
$M_R$ cause that the mixing matrix elements
$(K,K_{R})_{eN_1},(K,K_{R})_{eN_2}$,
which are a part of the unitary matrix U, change not only their phase
factors but absolute values, too.
So, firstly there is a real CP effect - phases in $({K,K_R})_{eN}$ matrix
elements change with changing $\alpha,\beta,\gamma$ and cause
different interferences among various diagrams. Secondly, different 
absolute values of mixing matrix elements are obtained that result also in
changes of the cross section magnitude.

In Fig.4 both these
effects are taken into account together. To find out the influence of CP phases only
we fix in Fig.5 absolute values of mixing matrix elements to be constant with
changing $\alpha,\beta,\gamma$.
\newpage
\ \\

\vspace{6.5 cm}
\begin{figure}[h]
\includegraphics{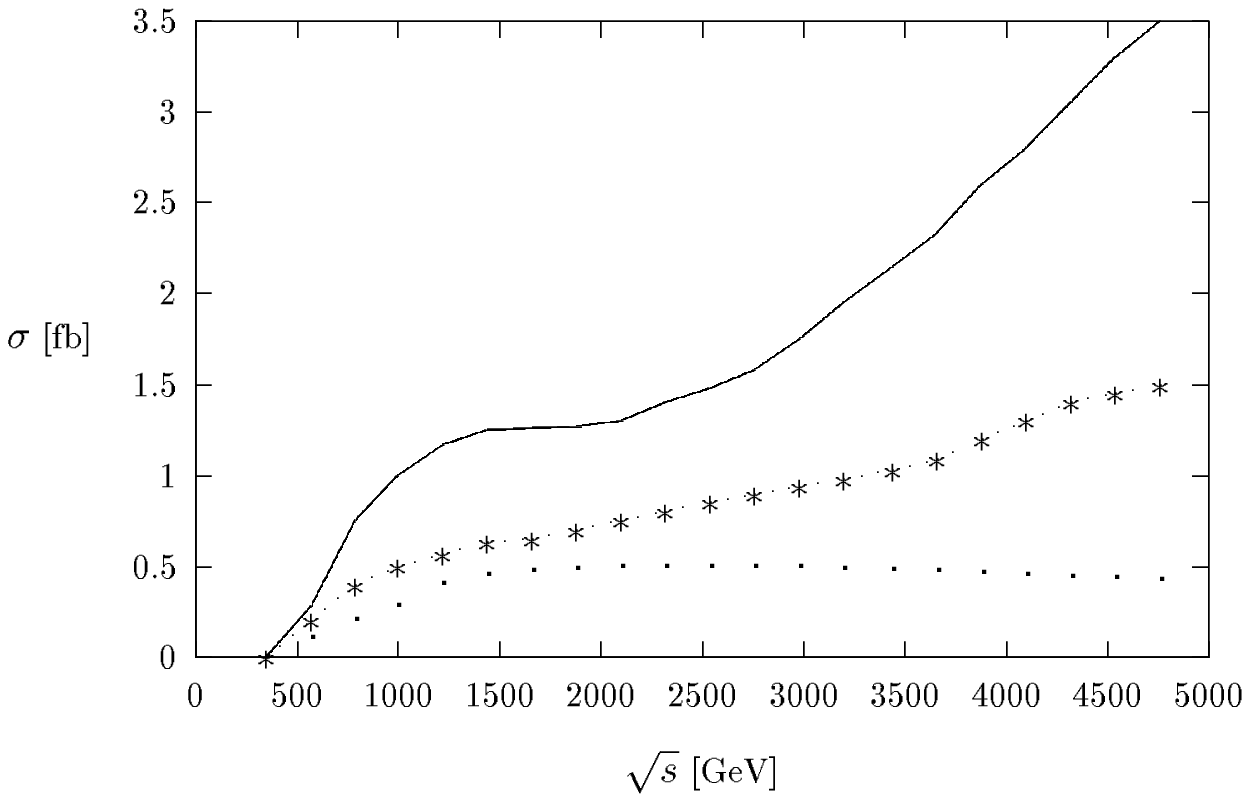}
\end{figure}
\baselineskip 5 mm

{\footnotesize Fig.4. CP and mixing matrix effects for the $e^-e^+
\rightarrow N_1(150)N_2(200)$ production process. Solid line is for
$\alpha=\beta=\gamma=0$, dotted line is for $\alpha=\pi,\beta=\gamma=0$ and
the third line (with asterisks) is for $\alpha=2.0,\beta=\gamma=0$ phases.}
\\

\baselineskip 7 mm

We can see that the influence of the CP
interference is very large. The cross section for production of two neutrinos
with opposite CP parities can be several times
larger then the cross section for production of the same CP parity neutrinos.
The cross sections for the real CP breaking case are placed between two CP
conserving situations. 

We can see from Fig.4 and Fig.5  that the total cross section for production
of two different heavy neutrinos equals a few femtobarns  for $\sqrt{s} \leq
0(1)$ TeV. The cross section for production of two identical heavy neutrinos
could be utmost $\sim$ 30 times larger (compare ${(K_R)}_{1e}$ and
${(K_R)}_{2e}$ in Fig.5 caption). It could happen if the mixing angle
${(K_R)}_{Ne}$ have been maximal: ${(K_R)}_{Ne}\sim 1$. However, in this 
case CP effects disappear (see Eqs.(3.4)-(3.5)). 
These  cross sections
depend crucially on the mass of the additional gauge boson $W_2$ and are
negligible for the RHS model as a right-handed heavy neutrino couples
very weakly to a gauge boson $W^-$.

\vspace{6.5 cm}
\begin{figure}[h]
\includegraphics{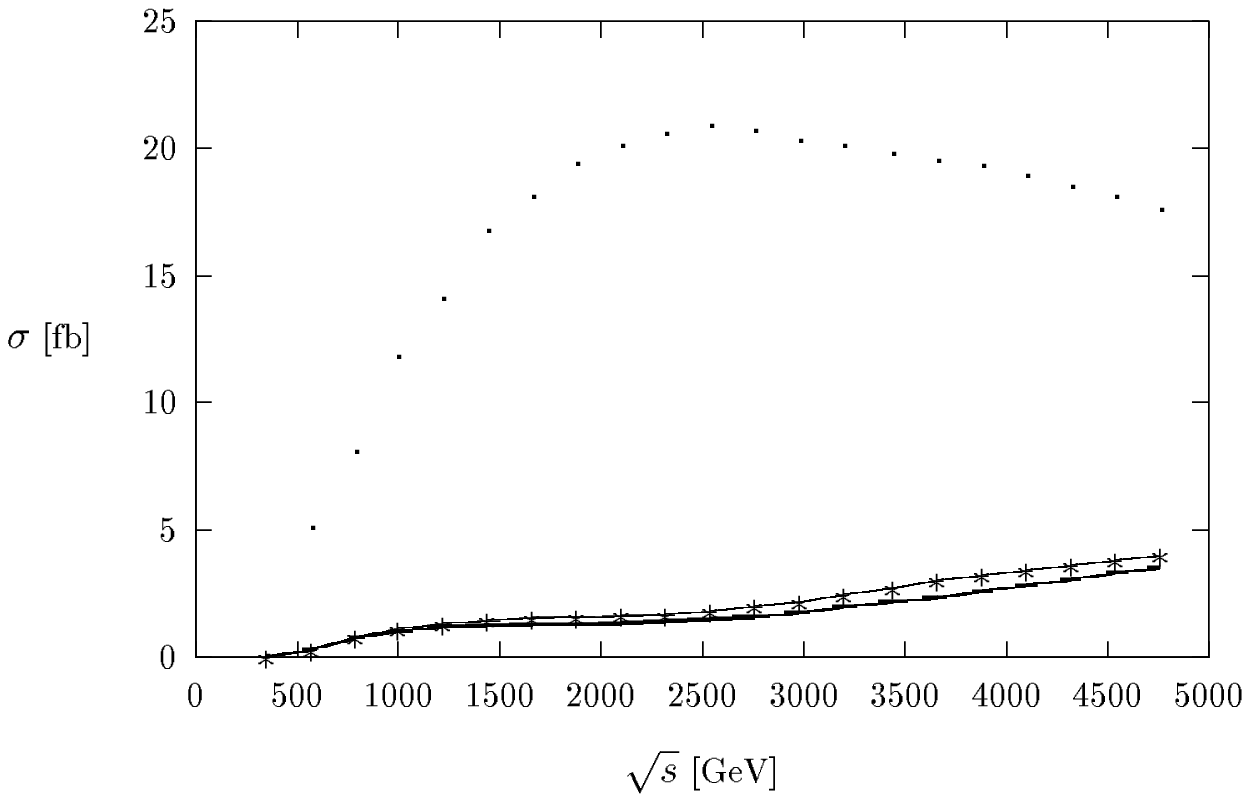}
\end{figure}

\baselineskip 5 mm

{ \footnotesize Fig.5.~The pure CP violation effect caused by phases
on the $e^-e^+ \rightarrow N_1(150)N_2(200)$ production process. Absolute
values of mixing matrix elements for all lines are the same as the ones for the solid line
in Fig.4.
$[{(K)}_{1e}=0.00535,{(K_R)}_{1e}=0.9819,K_{2e}=0.0058, \newline
{(K_R)}_{2e}=0.189]$.
The dotted (solid) line is for the opposite (the same) CP parity of neutrinos
(Eqs.(3.25) and (3.24)); line with asterisks is for $\alpha=2.0,\beta=\gamma=0$, the
same as in Fig.4.}
\\

\baselineskip 7 mm

All numerical results which have been shown till now in this Section have been
connected with the `see-saw' model. Because the mixing angles ${(K_R)}_{Ne}$
have been already of the order of 1 (e.g. maximal) we can expect that numerical
results for two heavy neutrino production process are of the same order for
the
`non-decoupling' model, too. Another situation will be, however, for a single 
heavy neutrino production process $e^-e^+ \rightarrow \nu N$ as a heavy
neutrino-electron mixing angle $K_{Ne}$ can be different for these models
(Section 2). The situation for
the `see-saw' model is summarized in Fig.6  where  the cross section as a
heavy neutrino mass function is shown. 

We can see that the values of the cross section are significant 
(range of few fb) only 
for small masses of a heavy neutrino.
Now, in the last subsection I would like to focus on `non-decoupling'
effects in the single heavy neutrino production process.

\vspace{7 cm}
\begin{figure}[h]
\includegraphics{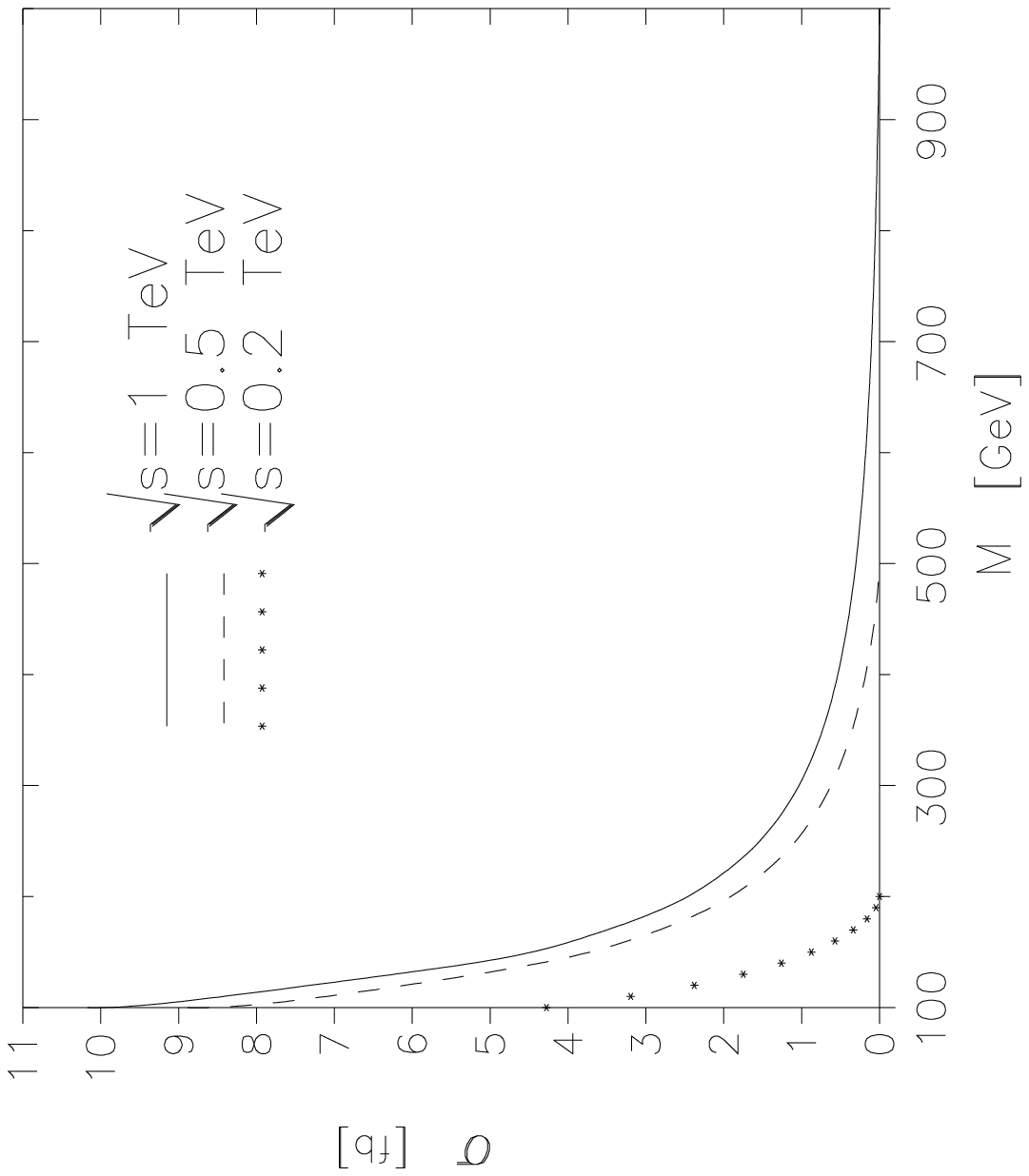}
\end{figure}

\baselineskip 5 mm

{\footnotesize Fig.6. The cross section for the $e^-e^+ \rightarrow
\nu N$ process in the framework of the classical `see-saw' models. Solid, dashed
lines and the line with stars are for 1 TeV, 500 GeV and 200 GeV CM energies of
future colliders, respectively. } \\

\baselineskip 7 mm

\subsection{`Non-decoupling' model and a heavy neutrino in the $e^-e^+$
collision}

As it has been  already shown in Section 2 the maximal mixing angle $K_{Ne}$ 
depends on the number of neutrinos. 

$\bullet\;n_R=1$ (or all heavy neutrinos with the same $\eta_{CP}$'s) \\

According to Eq.(2.23) mixing angle $K_{Ne}$ depend crucially on $\omega$.
Results for this case are  given in Table 3.  
\newpage
%
{\footnotesize
\begin{table}[t]
\begin{center}
\begin{tabular}{|c| c| c| c|} 
\cline{1-4}
\cline{1-4}
& \multicolumn{3}{|c|}{ }  \\ 
$M_N$ [GeV] & \multicolumn{3}{|c|}{ $\sigma^{total}_{max}$ [fb], $n_R=1$ }  \\ 
& \multicolumn{3}{|c|}{ }  \\ 
&&& \\
& $\sqrt{s}=0.5$ TeV & $\sqrt{s}=1$ TeV & $\sqrt{s}=2$ TeV \\ 
&&& \\
\cline{1-4}
\cline{1-4}
&&& \\
100 & 0.18 & 0.2 & 0.2 \\ 
&&& \\
150 & 0.25 & 0.3 & 0.3 \\
&&& \\
200 & 0.31 & 0.4 & 0.4 \\
&&& \\
300 & 0.34 & 0.6 & 0.6 \\
&&& \\
500 & - & 0.8 & 1.0 \\
&&& \\
700 & - & 0.7 & 1.3 \\
&&& \\
1000 & - & - & 1.6 \\
\cline{1-4}
\cline{1-4}
\end{tabular}
\vspace{ 0.5cm}
\end{center}
\end{table} }
\baselineskip 5 mm 
Table 3: {\footnotesize The total cross section $\sigma_{tot} \left( e^+e^- \rightarrow \nu N
\right) $ in $n_R=1$ case (see Eq.(2.23) with 
$\omega^2=2 \cdot 10^{-5}\;TeV^{-1}$)
for various heavy neutrino masses and three different total energies $\sqrt{s}=
0.5,1,2$ TeV. }

\baselineskip 7 mm
\vspace{1 cm}
As the maximal value of the $K_{Ne}$ parameter
is proportional to $M_N$ the cross section is an increasing function of a heavy
neutrino mass with exception at the end of the phase space $M_N \rightarrow \sqrt{s}$.
Results are miserable\footnote{The last results obtained for $\omega$
\cite{hmraid} can increase these values by a factor $\sim 40$ and then
$n_R=1$ case can focus some interest, too.}.

\setlength{\textwidth}{14 cm}
\setlength{\textheight}{19 cm}
\setlength{\topmargin}{1 mm}
\newpage
$\bullet\;n_R=2$ \\

Taking into account Eqs.(2.28-2.29) in Fig.7 the 
cross section for $e^+e^- \rightarrow N\nu $ process
as a function of the lighter neutrino's mass for different values of
$A=\frac{M_2}{M_1}$ factor and for  $\sqrt{s}=1$~TeV is depicted. 
There is space for
large $\sigma$ but only for very small mass differences $(M_1
\simeq M_2)$ - see Eq.(2.31). I take maximal $\kappa^2$ that equals to 0.0054
(Eq.(2.16)).

\vspace{6 cm}
\begin{figure}[h]
\vspace{.1 cm}
\includegraphics{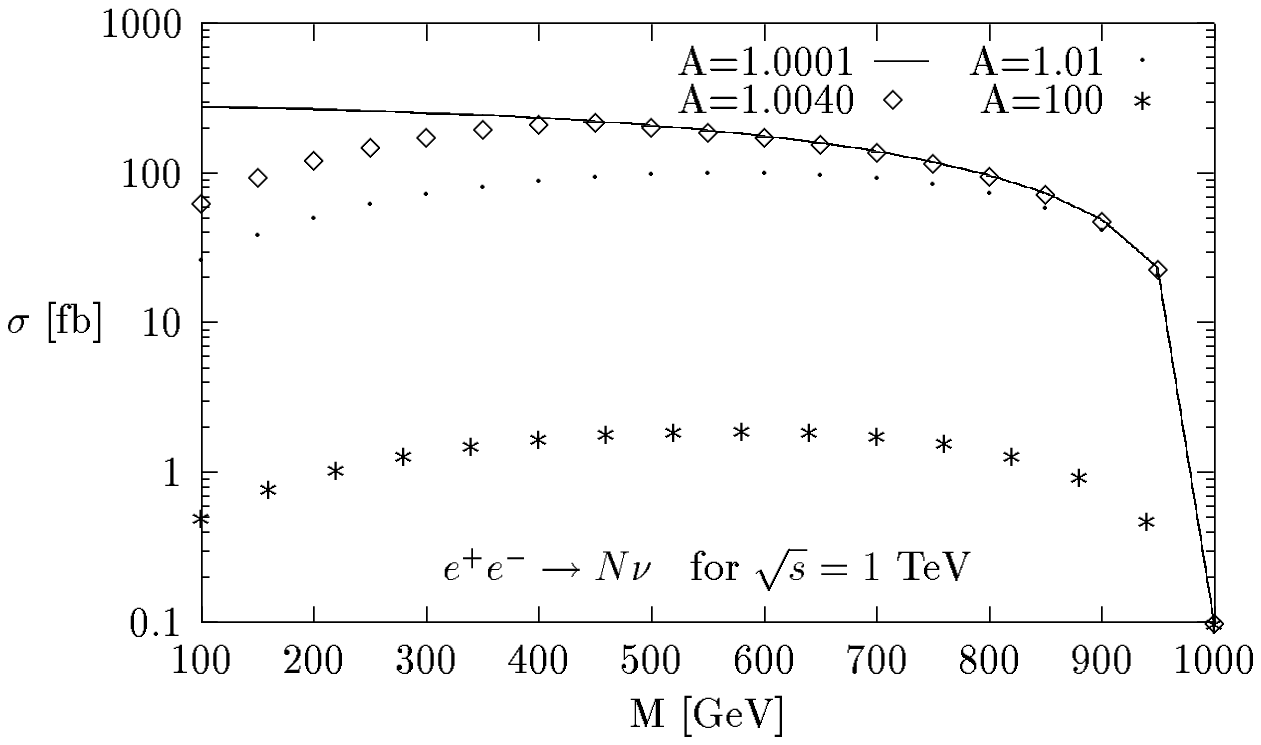}
\vspace{0.1 cm}
\end{figure}

\baselineskip 5 mm
{\footnotesize Fig.7. The cross section for the $e^+e^- \rightarrow N\nu$ 
process as a function of a heavy
neutrino mass $M_1=M$ for $\sqrt{s}=1$ TeV in the models with two heavy
neutrinos ($n_R=2$) for
different values of $A=\frac{M_2}{M_1}$ (solid line with $A=1.0001$,
`$\diamond$' line with
$A=1.004$, dots line with A=1.01 and `$\ast$' line with A=100). Only for a very
small mass difference 
$A \sim 1$ the existing experimental data leave the chance that the cross section
is large, e.g.\
$\sigma_{max}(M=100\;GeV) =275$~fb. If $M_2\gg M_1$ then the cross section must
be small, e.g.\ for
$A=100$, $\sigma_{max}(M=100 \rm\;GeV) \simeq0.5$ fb. }
\\

\baselineskip 7 mm 

$\bullet\;n_R=3$ \\

Results for this case are gathered in Fig.8. We have three heavy neutrinos
with masses $M_1=M, M_2=AM, M_3=BM$ and $\eta_{CP}(N_1)=\eta_{CP}(N_2)=
-\eta_{CP}(N_3)=+i$ (see Section 2). Then the maximal mixing angle of the
lightest heavy neutrino is $(A \rightarrow \infty , B \rightarrow 1)$
$K_{N_1e}^2=\frac{\kappa^2}{2}$ (Eq.(2.36)). The lower of two lines for
each of CM energies depicted in Fig.8 realizes this case. The upper line is
for the third of heavy neutrinos when its mixing angle is maximal, too.
It can happen when $B=\frac{M_3}{M_1}$
is as big as possible (Eq.(2.36)), so I take $M_1=100$ GeV and parameterize 
the mixing angle of the third heavy neutrino as follows
\begin{equation}
{\left( K_{N_3e} \right)}^2=\frac{\left(\frac{M}{100\;GeV} \right)\kappa^2}
{1+\left( \frac{M}{100\;GeV} \right)}.
\end{equation}
\vspace{7 cm}
\begin{figure}[h]
\vspace{.1 cm}
\includegraphics{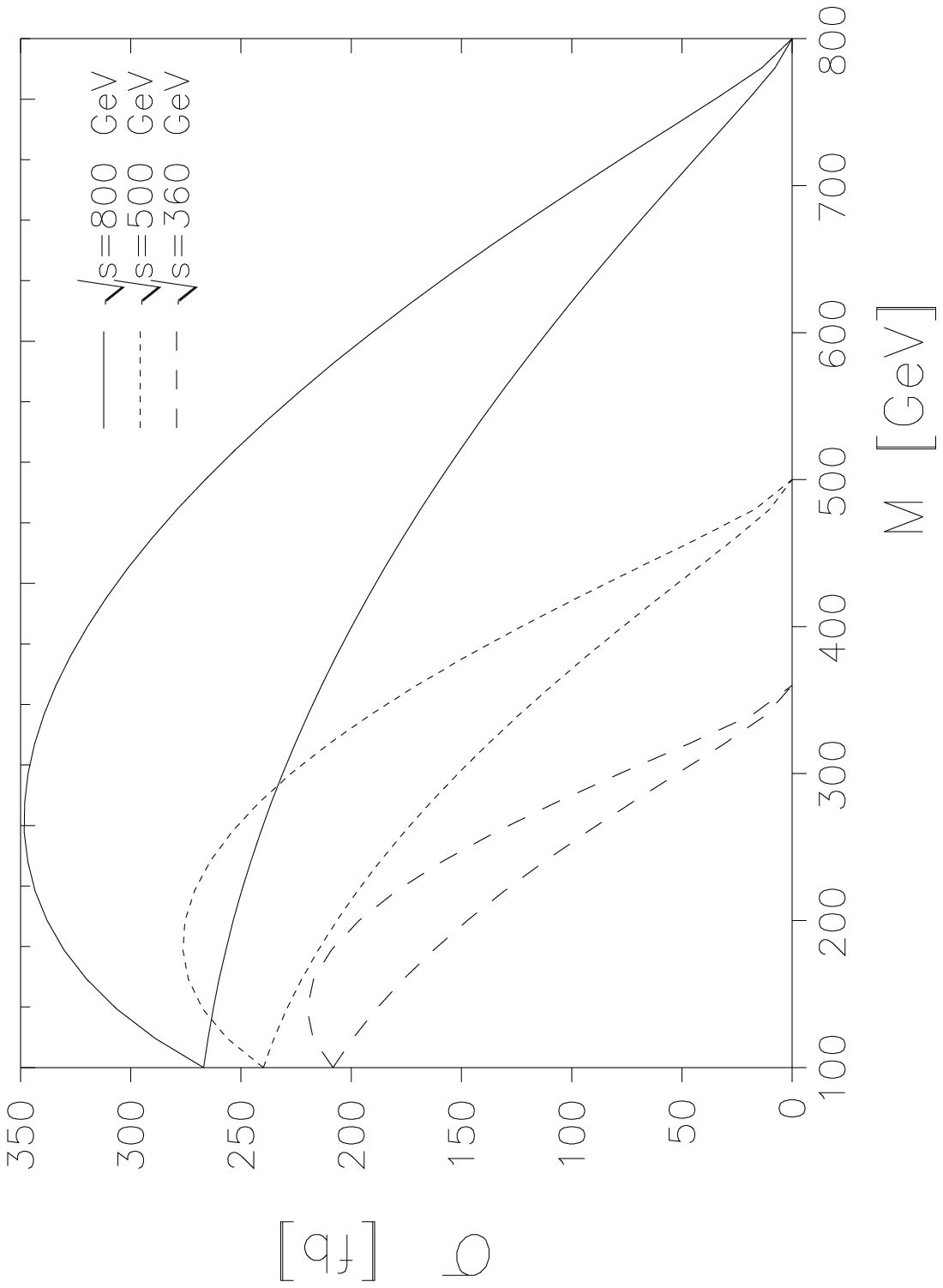}
\vspace{0.1 cm}
\end{figure}

\baselineskip 5 mm
{\footnotesize Fig.8 Production of a single heavy neutrino $(e^-e^+ \rightarrow
\nu N)$ with maximal possible mixing angle $K_{Ne}$ for different CM energies:
$\sqrt{s}=360,500,800$ GeV as a function of heavy neutrino mass. The lower line in
each pair is for the lightest of heavy neutrinos, the upper is for the third of heavy
neutrinos with mass M and the mixing angle given by Eq.(3.26).}
\\

\baselineskip 7 mm
As it was already mentioned the results for the LR and the RHS models are practically
the same for the single heavy neutrino production (see Table 1). 
That is why the results
given in Fig.8 can be assumed to describe these two models altogether.
We can see that the results for the `non-decoupling' model are more optimistic
than for the `see-saw' one and it is worth to study what we can find in reality
in a collision \cite{pr9}. Heavy neutrinos which I consider are unstable.
Possible decay channels are $N \rightarrow  W^{\pm}l^{\mp}$,
$N \rightarrow  Z \nu_l$ and $N \rightarrow  H \nu_l$. Two first are always
opened for $M_N \geq 100$ GeV, the last one depends on Higgs particle's mass.

The total decay width equals to
\begin{equation}
\Gamma_N=\sum_l \left( {2\Gamma(N \rightarrow l^+W^-)+ \Gamma(N \rightarrow \nu_l Z)
+\Gamma(N \rightarrow \nu_l H)\Theta(m_N-m_H)} \right)
\end{equation}
where
\begin{equation}
\sum_l \Gamma(N \rightarrow l^+W^-) \propto \sum\limits_{l=e,\mu,\tau}
\mid K_{Nl} \mid^2 \simeq \mid K_{Ne} \mid^2,
\end{equation}
\begin{equation}
\sum_l \Gamma(N \rightarrow \nu_l H),
\sum_l \Gamma(N \rightarrow \nu_l Z) \propto \sum_l
\mid \Omega_{N\nu_l} \mid^2 \simeq \sum_l \mid K_{Nl} \mid^2 \simeq
\mid K_{Ne} \mid^2.
\end{equation}
In the approximations made in Eqs.(3.28) and (3.29) it is assumed that in each column 
of K matrix ($l=e,\mu,\tau$)
$$(K_{\nu_el}, K_{\nu_{\mu}l}, K_{\nu_{\tau}l}, K_{N_1l}, K_{N_2l}, ...)^T$$
only one coupling
between heavy neutrinos and lepton is visible $K_{N_il} \simeq K_{Ne}$. 
All other couplings are very small and are neglected. 

If we also assume 
lepton universality $(K_{\nu_ll} \simeq 1)$
then also the production cross section can be parameterized by only one 
mixing angle
\begin{equation}
\sigma_{tot}=\sum\limits_{i=e,\mu,\tau} \sigma( e^+e^- \rightarrow \nu_i N),
\end{equation}
and
\begin{eqnarray}
\sigma_{tot} & \propto & \mid K_{Ne} \mid^2 \left( \mid K_{\nu_e e}
\mid^2+\mid K_{\nu_{\mu}} \mid^2 + \mid K_{\nu_{\tau}} \mid^2 \right) 
\nonumber \\
&=& \mid K_{Ne} \mid^2 ( 1- \sum_N \mid K_{Ne} \mid^2 )^2
\simeq \mid K_{Ne} \mid^2.
\end{eqnarray}

In this way we have only one parameter $K_{Ne}$ which is important in the 
neutrino production-decay process.

So, let's discuss the angular distribution for the final electron
(positron) in the process
\arraycolsep0.5mm
\begin{equation}
e^+e^- \rightarrow  
\begin{array}[t]{ll}
\nu & N \qquad  \\
& \hookrightarrow  e^{\pm}W^{\mp}
\end{array}
\end{equation}
where for $N$ I take the lightest of heavy neutrinos.

Numerical results are gathered on the next three Figures.

\vspace{7 cm}
\begin{figure}[h]
\vspace{.1 cm}
\includegraphics{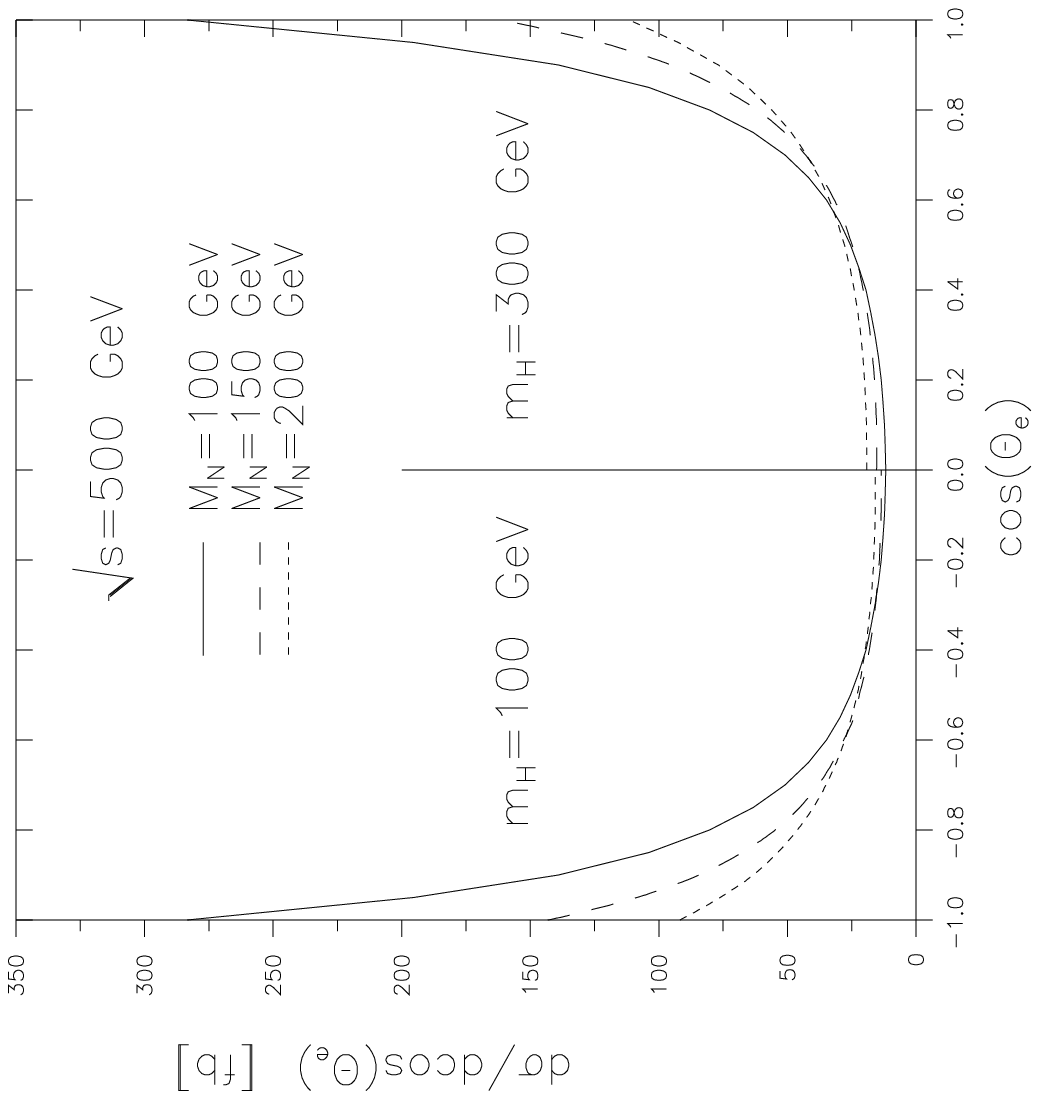}
\vspace{0.1 cm}
\end{figure}

\baselineskip 5 mm
{\footnotesize Fig.9. Distribution of the final electron from a heavy neutrino decay for 
$\sqrt{s}=500$ GeV collider with $M_N=100$ GeV (solid line), $M_N=150$ GeV
(long-dashed line) and $M_N=200$ GeV (short-dashed line). Left
half of the Figure gives results for $m_H=100$ GeV, right half of the Figure 
for $m_H=300$ GeV (Higgs decay channel is closed).}
\\

\baselineskip 7 mm

In Fig.9 I present the angular distribution for the final electron
$e^-e^+ \rightarrow \nu (N \rightarrow e^-W^+)$ for
various masses of heavy neutrino $M_N=100,150$ and 200 GeV calculated for
the maximal possible $\mid K_{Ne} \mid^2 \simeq \frac{\kappa^2}{2}$ ($\kappa^2
=0.0054$).
Results are given for the NLC
with $\sqrt{s}=500$ GeV. This distribution has forward-backward
symmetry. To show the influence of Higgs particle I include results for
$m_H=100$ GeV on the
left half of the Figure  $(-1 \leq \cos{\Theta_e} \leq 0)$  
and on the right side $(0 \leq \cos{\Theta_e} \leq 1)$
for $m_H=300$ GeV. For a higher Higgs mass the total width $\Gamma_N$ is
smaller due to the greater value of the branching ratio for the $N \rightarrow lW$
decay and the cross section $\frac{d\sigma}{d \cos{\Theta_e}}$ is larger. 
Numerically, Higgs has no influence on the cross section for $M_N=100$ GeV
(for $m_H=100 \div 300$ GeV $N \rightarrow \nu H$ decay channel is closed)
and the influence of the Higgs particle is
approximately equal to 10 \%, 15\% for $M_N=150,200$ GeV, respectively 
(only the decay mode with $m_H=100$ GeV is opened in this case).

\vspace{7 cm}
\begin{figure}[h]
\vspace{.1 cm}
\includegraphics{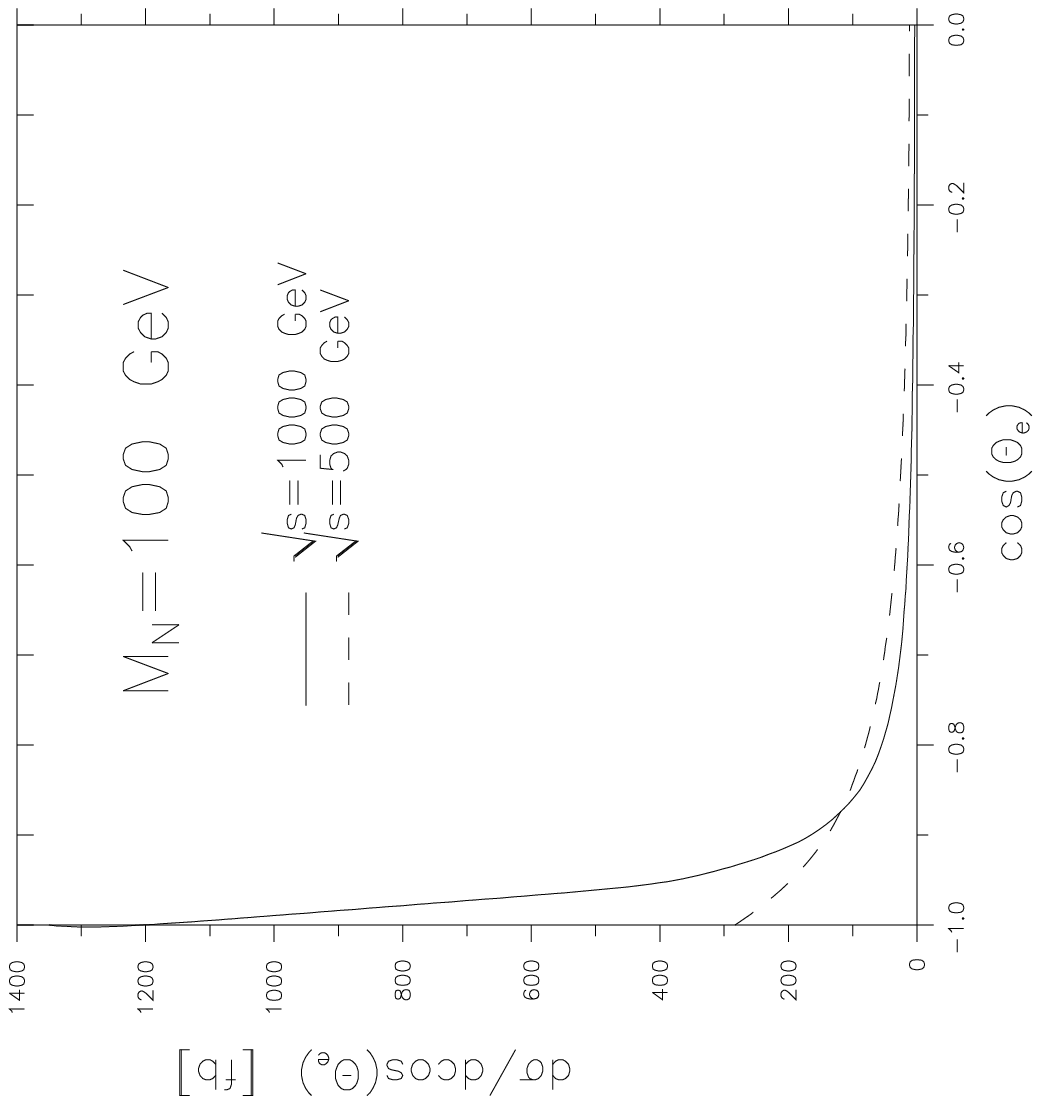}
\vspace{0.1 cm}
\end{figure}

\baselineskip 5 mm
{\footnotesize Fig.10 Backward distribution of the final electron coming from 
a heavy neutrino
decay ($M_N=100$ GeV) for two different energies: $\sqrt{s}=500$ GeV
(dashed line) and $\sqrt{s}=1000$ GeV (solid line).}
\\

\baselineskip 7 mm

For higher energies the final electron distribution is more peaked
in the forward-backward direction $( \cos{\Theta_e}=\pm1)$. This is the result
of $W^{\pm}$ exchange in t and u channels. As an
example I have compared the final electron distribution from the decay of a 
heavy neutrino with mass $M_N=100$ GeV for $\sqrt{s}=500$ GeV and 
$\sqrt{s}=1000$ GeV energies (Fig.10).

Finally in Fig.11 I present the angular distribution $\frac{d\sigma}
{d\cos{\Theta_e}}$ for various masses of heavy neutrino $M_N=100,300$ and
500 GeV ($m_H=100$ GeV). The cross section becomes higher and more
peaked in the forward-backward direction for a smaller mass of heavy neutrinos.
The effect of growing $\frac{d\sigma}{d \cos{\Theta_e}}$ is the result
of increasing $BR(N \rightarrow lW)$ and increasing $\sigma_{tot}
(e^+e^- \rightarrow \nu N)$ for smaller $M_N$. The effect
of slope reduction with $M_N$ mass in the forward-backward direction is also
kinematically understandable.

\vspace{7 cm}
\begin{figure}[h]
\vspace{.1 cm}
\includegraphics{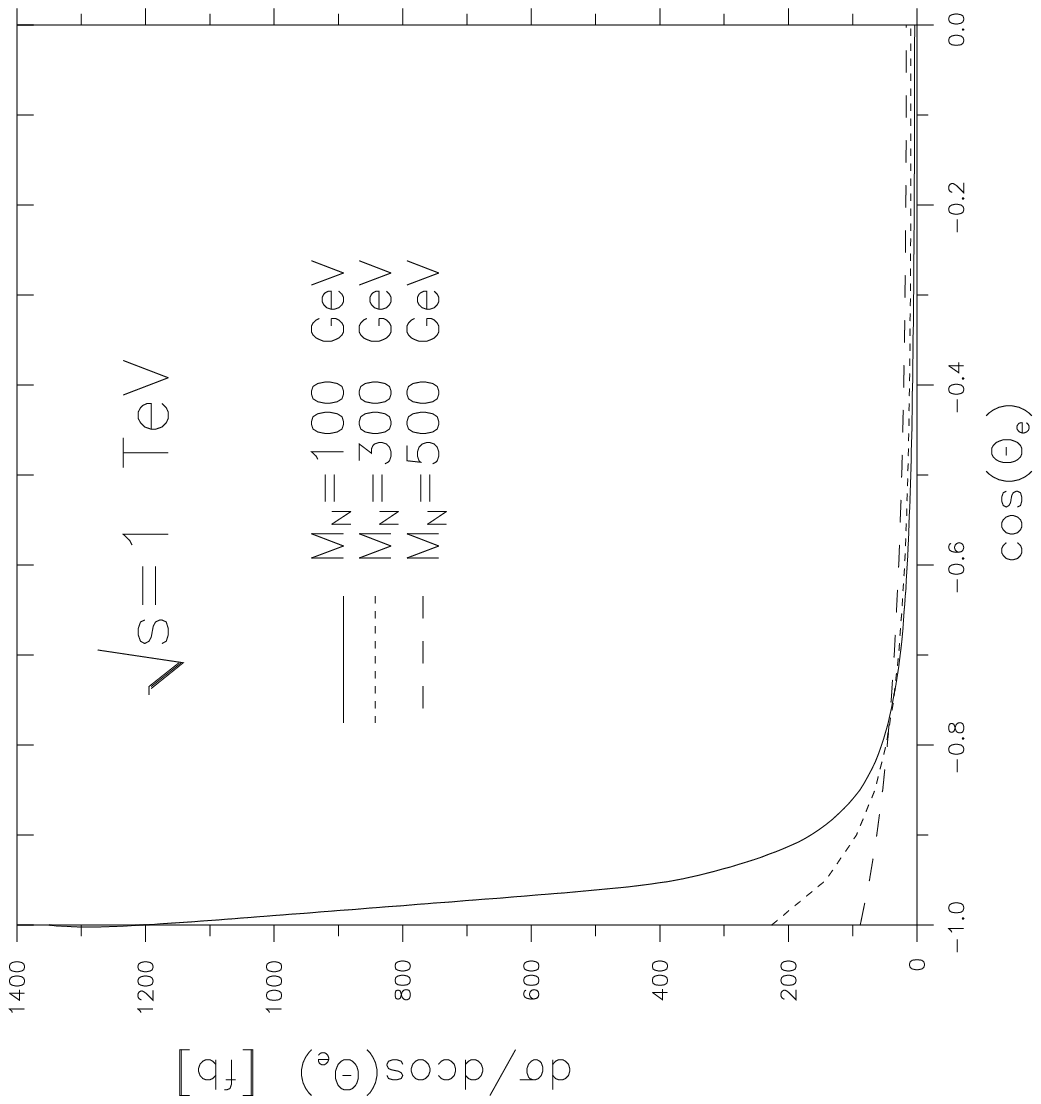}
\vspace{0.1 cm}
\end{figure}

\baselineskip 5 mm
{\footnotesize Fig.11. Backward distribution of the final electron coming from 
a heavy neutrino
decay with mass $M_N=100$ GeV (solid line), $M_N=300$ GeV (short-dashed line)
and $M_N=500$ GeV (long-dashed line) for $\sqrt{s}=1$ TeV.}
\\

\baselineskip 7 mm
Backward distribution of the electron coming from a heavy neutrino decay
gives a chance for a heavy neutrino detection.
The main background for this process is the production of $W^+W^-$ pair 
with the decay
$W^{-} \rightarrow e^{-} \nu$. 
The distribution of the electron coming from the heavy neutrino decay 
(N) and from W's decay by $e^+e^- \rightarrow
W^+W^-$ process differs very much in the forward-backward direction.  
For a high energy ($\sqrt{s}>0.5$ TeV) the angular distribution of electrons
coming from the $W^-$ decay is peaked in the forward direction. On the
contrary,
the $e^-$ coming from the N decay will travel equally well both in forward 
and backward direction.
\newpage







\setcounter{equation}{0}
\renewcommand{\theequation}{4.\arabic{equation}}
\section{Indirect heavy neutrinos detection in $e^-e^-$ collision:
$e^-e^- \rightarrow W^-W^-$ process}
The $e^-e^- \rightarrow W^-W^-$ process was firstly proposed in 1982
(\cite{rizzo}, see also \cite{eeww}) as one of tests for the lepton 
number violation. As it can be seen from 
Fig.12 this process is sensible to all neutrinos, light and heavy. In the
RHS model only t and u channels are present while in the LR model doubly charged
Higgs particles are important, too. They introduce resonances and are needed
for proper high-energy behaviour of the cross section
\cite{rizzo},\cite{pr5}. 

\vspace{5 cm}
\begin{figure}[h]
\includegraphics{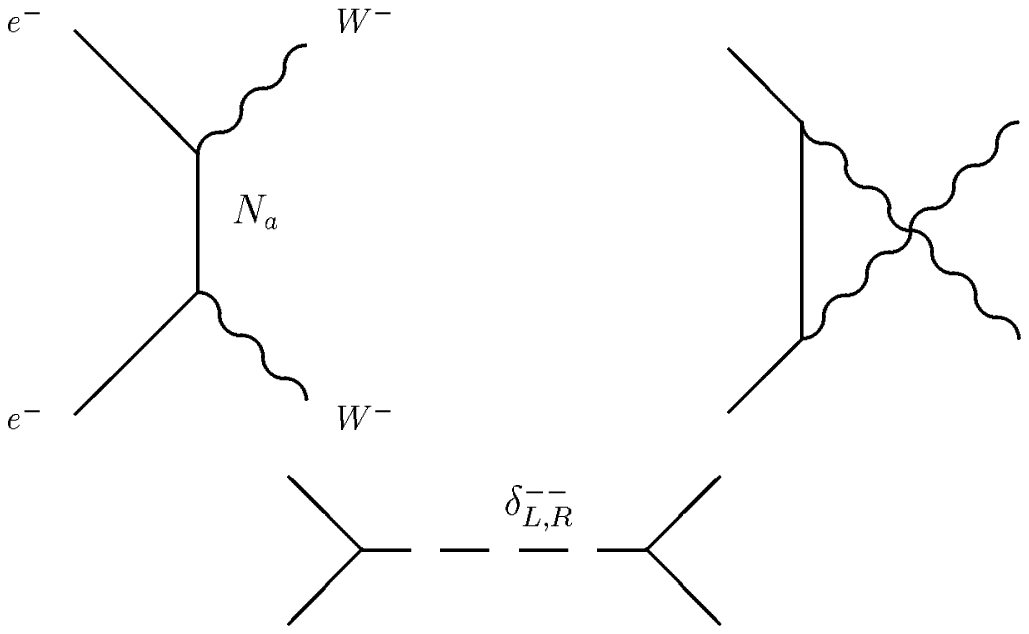}
\end{figure}

\baselineskip 5 mm
{\footnotesize Fig.12. The tree level Feynman diagrams which contribute to
the $e^-e^- \rightarrow W^-W^-$ process. In the LR model all three channels
are present, while in the RHS model s-channel is absent.} \\

\baselineskip 7 mm
I give up here a detailed analysis of the additional processes that can appear 
in the LR model
($e^-e^- \rightarrow W^-_1W^-_2, e^-e^- \rightarrow W^-_2W^-_2$) and I will
focus on SM gauge bosons production. Because of kinematical reasons these 
additional processes can be
of special importance not before a collider with $\sqrt{s} \simeq 1$ TeV energy
will appear. For such energies, especially when
$W_2^-$ would be not too heavy (let's say $M_{W_2} \simeq 600$ GeV that is
close to the present experimental limit \cite{wr},\cite{wrhad}) the $W_2^-$
would be easily discovered unless the energy of collider would be too close
to the 
threshold \cite{pr5},\cite{maal}.

Without heavy neutrinos the process $e^-e^- \rightarrow W^-W^-$  is
negligible
small. The reason is that light (left-handed) neutrinos are (almost) massless
and such conditions cause that Majorana neutrinos are not distinguishable
from Dirac neutrinos (actually they decouple to Weyl spinors) \cite{weyl}
and the lepton number is conserved (see \cite{bil} for details). 
Contribution of a left-handed doubly charged Higgs particle in s-channel is
negligible, too \cite{pr10}. 
Apart from heavy neutrinos, $\delta_R^{--}$ right-handed resonance is
important for this process, too.
This case will be discussed later.

I give up here writing down the helicity amplitudes for this process. They
can be found in \cite{pr6},\cite{pr5}. Two facts are important. Firstly, as
all heavy neutrinos are exchanged, the amplitude is a sum of all of them.
Secondly, we have checked that for the $e^-e^- \rightarrow W^-W^-$ process the
helicity amplitudes with purely left-handed or purely right-handed
electrons, proportional to the heavy neutrino
masses, are dominant \cite{pr5}. These helicity amplitudes include either a 
square of the $K$ mixing
matrix elements ($e_L^-e_L^-$ collision) or a square of the $K_R$ ones 
($e_R^-e_R^-$ collision). That is why these two facts cause that interferences among contributions from
different neutrinos can appear if some $K$ or $K_R$ elements are complex. 
Let's discuss these effects more carefully now.

\subsection{CP violation effects in the $e^-e^- \rightarrow W^-W^-$
process (`see-saw' model)}

Similarly to the discussion of CP effects in the $e^-e^+ \rightarrow
\nu N$ process (Section 3) we choose the neutrino mass matrix $M_R$ in the form 
\begin{equation}
M_R=\left( \matrix{ e^{i\alpha}M & 10 &20 \cr
                10 & 2e^{i\beta}M & 10 \cr
                20 & 10 & 3e^{i\gamma}M } \right).
\end{equation}

Matrix $m_D$ is the same as in Eq.(3.22).

Fig.13 gives results for the LR model where the above parameterization 
was used. 

\vspace{7.5 cm}
\begin{figure}[h]
\includegraphics{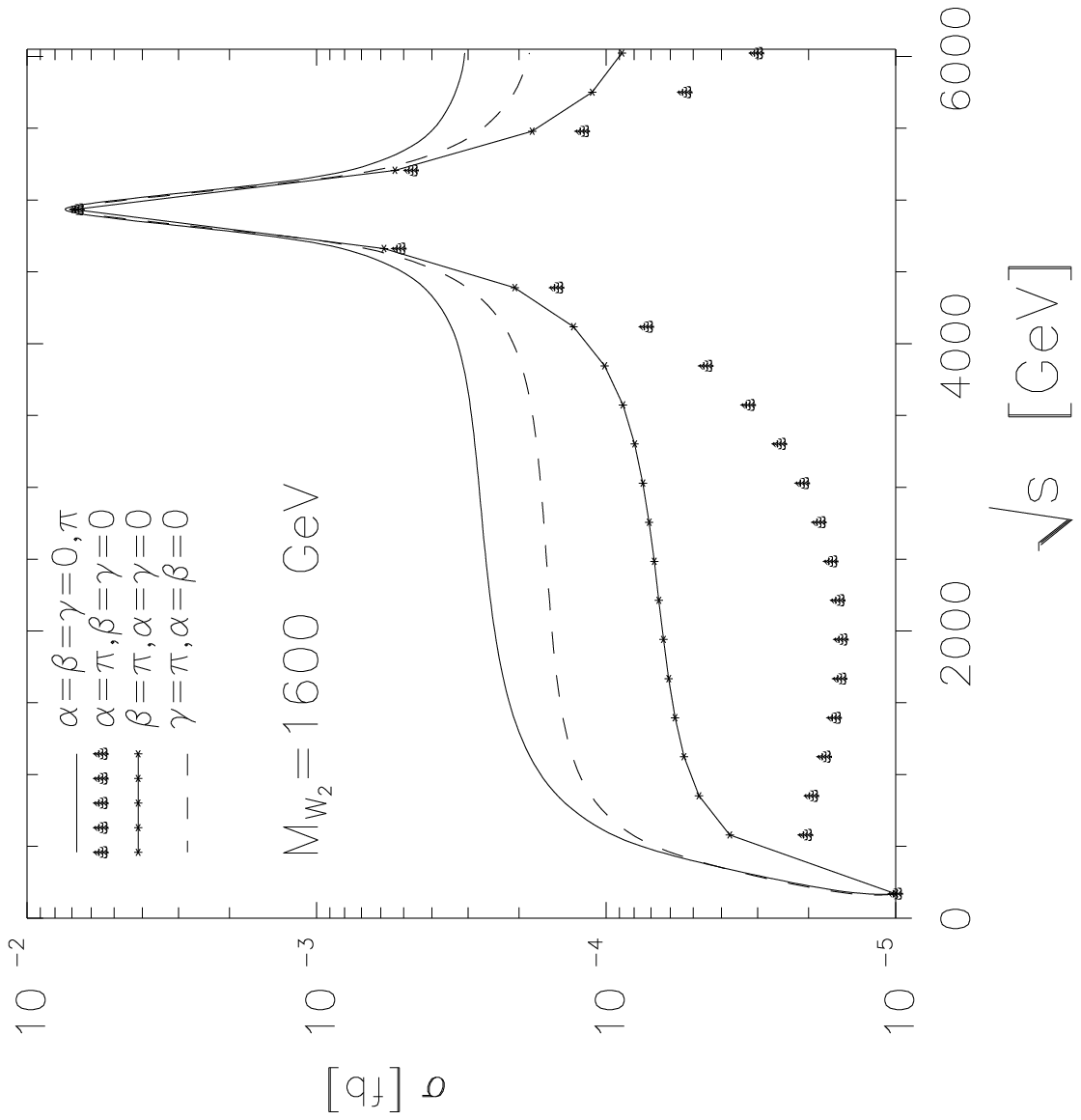}
\end{figure}

\baselineskip 5 mm

{\footnotesize Fig.13. Process $e^-e^- \rightarrow W^-_1W^-_1$ (LR model)
for $M_1=$200 GeV, $M_2=$400 GeV, $M_3=$600 GeV (Eq.(4.1)) and $M_{W_2}=$1600
GeV when CP parity is conserved in the lepton sector. } \\

\baselineskip 7 mm

The shape of these lines can be understood in the
following way.

If all diagonal elements $M_i$ (i=1,2,3) of the matrix $M_R$ (Eq.(4.1))
 are real and positive $\left(
\alpha =\beta=\gamma=0 \right)$ then the eigenvalues of the neutrino mass
matrix are also positive and the CP symmetry is conserved if the CP parities of
heavy neutrinos are the same and equal
\begin{equation}
{\eta}_{CP}{\left( N_4 \right)}={\eta}_{CP}{\left( N_5 \right)}=
{\eta}_{CP}{\left( N_6 \right)}=+i .
\end{equation}
The same happens when all masses $M_1,M_2,M_3$
are real, negative $( \alpha=\beta= \newline \gamma =\pi )$. Then CP is 
also conserved if CP parities of neutrinos are negative, imaginary 
\begin{equation}
{\eta}_{CP}{\left( N_4 \right)}={\eta}_{CP}{\left( N_5 \right)}=
{\eta}_{CP}{\left( N_6 \right)}=-i .
\end{equation}

In both cases above the mixing matrix elements $\left( K,K_{R} \right)_{ei}$,
i=4,5,6 are either pure real $(\alpha=\beta=\gamma=0)$ or pure imaginary
$(\alpha=\beta=\gamma=\pi)$. 
As it was already mentioned the dominant
helicity amplitudes include either a square of the 
$K$ mixing
matrix elements ($e_L^-e_L^-$ collision) or a square of the $K_R$ ones 
($e_R^-e_R^-$ collision)
and a summation must be carried out over all exchanged heavy neutrinos.
That is why constructive contribution from all heavy
neutrinos is present for $\alpha=\beta=\gamma=0,\pi$ (solid line on Fig.13).
In the case of mixing CP parities the CP symmetry is also conserved but the destructive
interference between contributions from various neutrinos causes that the
cross section decreases. In this case some of $\left( K,K_{R} \right)_{ei}$,
i=4,5,6 matrix elements are real and some are imaginary.

To obtain these CP effects several $K$ or $K_R$ matrix elements
must interfere in the same helicity amplitude. The structure of the chosen
neutrino mass 
matrix causes that $K_{en}$ n=4,5,6 
matrix elements are of the similar order 
\begin{equation}
\mid K_{e4} \mid \simeq \frac{1}{M_4} >
\mid K_{e5} \mid \simeq \frac{1}{M_5} >
\mid K_{e6} \mid \simeq \frac{1}{M_6} 
\end{equation}
but only one suitable element of the $K_R$ matrix is large (see Eq.(2.11))
\begin{equation}
\mid K_R \mid_{e4} \sim 1 >> \mid \left( K_R \right) _{ei} \mid \;\;\;i=5,6.
\end{equation}
This property causes that only if the $K$ matrix elements contribute
to the cross section in the visible way ($e_L^-e_L^-$ collision), 
the CP breaking is seen.
It is just the case for two light gauge bosons $W_1^-W_1^-$ production where
$W^-_1$ couples predominantly with left-handed electrons (Eq.(B.21)).
If the contribution with $e_R^-e^-_R$ helicity amplitude becomes
important, then the CP symmetry effect decreases. That is why the
effect is visible only outside the $\delta_R^{--}$ resonance region where only 
one right-handed mixing matrix $K_R$ gives essential contribution (Eq.(4.5))
and the interference has no importance.
It also means
that the CP symmetry effect is more visible for the larger $M_{W_2}$, when
$\cos{\xi} \rightarrow 1$ (Eq.(B.21)). 

If the CP symmetry is violated (phases
$\alpha,\;\beta,\;\gamma \neq 0,\pi$), the cross sections lie
between two limiting lines in Fig.13. 

The effects of the CP symmetry in the RHS model  
is depicted in Fig.14. For the same energy with the same mass matrix 
parameterization the effects are practically the same as those
in the L-R model. 


\vspace{7.5 cm}
\begin{figure}[h]
\includegraphics{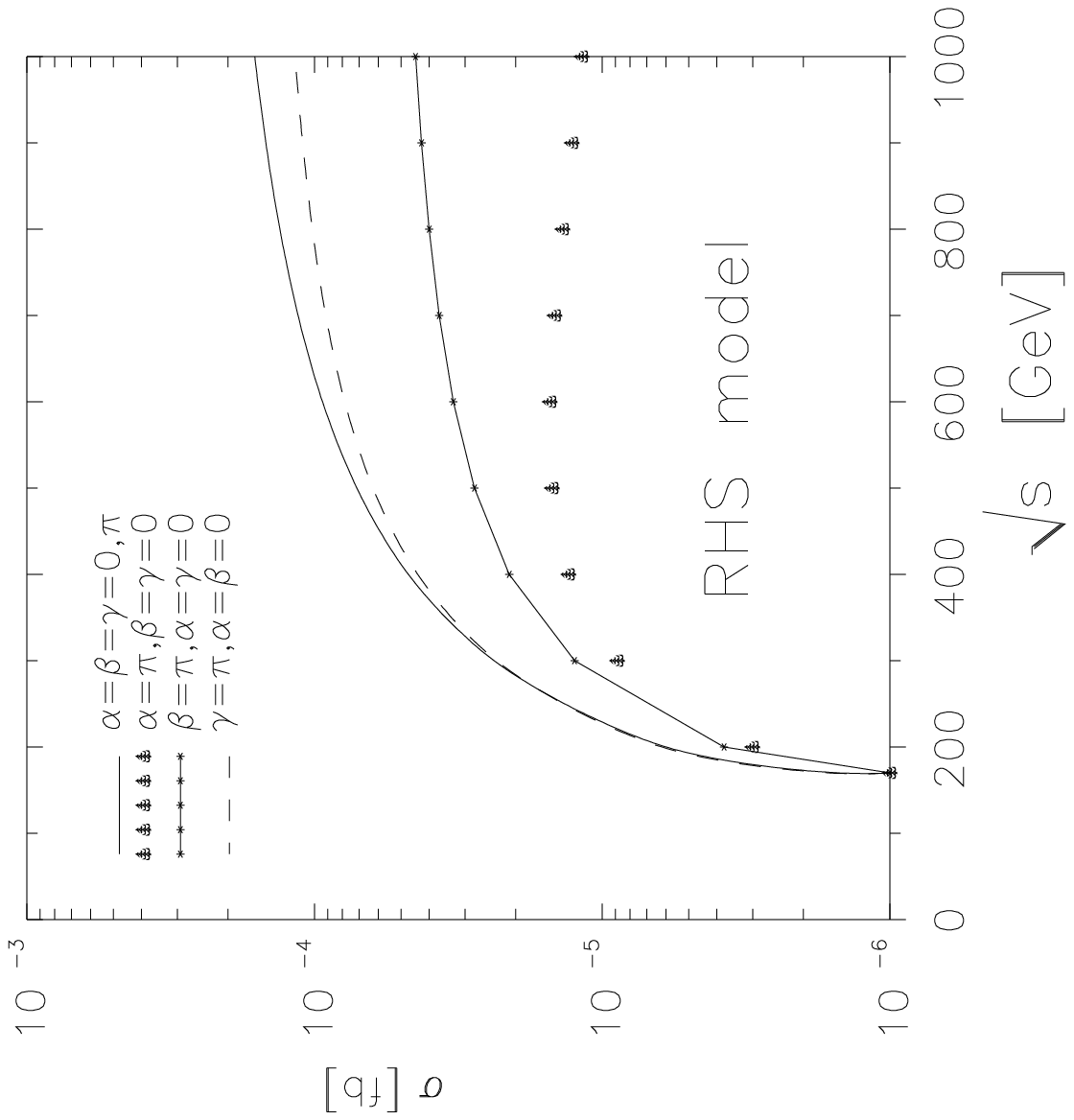}
\end{figure}

\baselineskip 5 mm
{\footnotesize Fig.14. Process $e^-e^- \rightarrow W^-_1W^-_1$ (RHS model)
for $M_1=$200 GeV, $M_2=$400 GeV, $M_3=$600 GeV 
when CP parity is conserved in the lepton sector. } 

\baselineskip 7 mm

We can see that although CP effects are quite interesting the cross sections
are very small. It is shown in Fig.15 where the
cross section as the function of heavy neutrino mass is carried out for a
classical `see-saw' model ($K_{Ne} \sim 1/M_N$). 

The signal from the $e^-e^- \rightarrow W^-W^-$ process is so clean (the
only important SM background comes from the $e^-e^- \rightarrow
W^-W^-\nu_e\nu_e$ process \cite{kol} and can be suppressed by appropriate
kinematical cuts) that as small as $\sigma=0.1$ fb cross section is enough
for possible process discovery \cite{kol1}.\\

\vspace{7 cm}
\begin{figure}[h]
\includegraphics{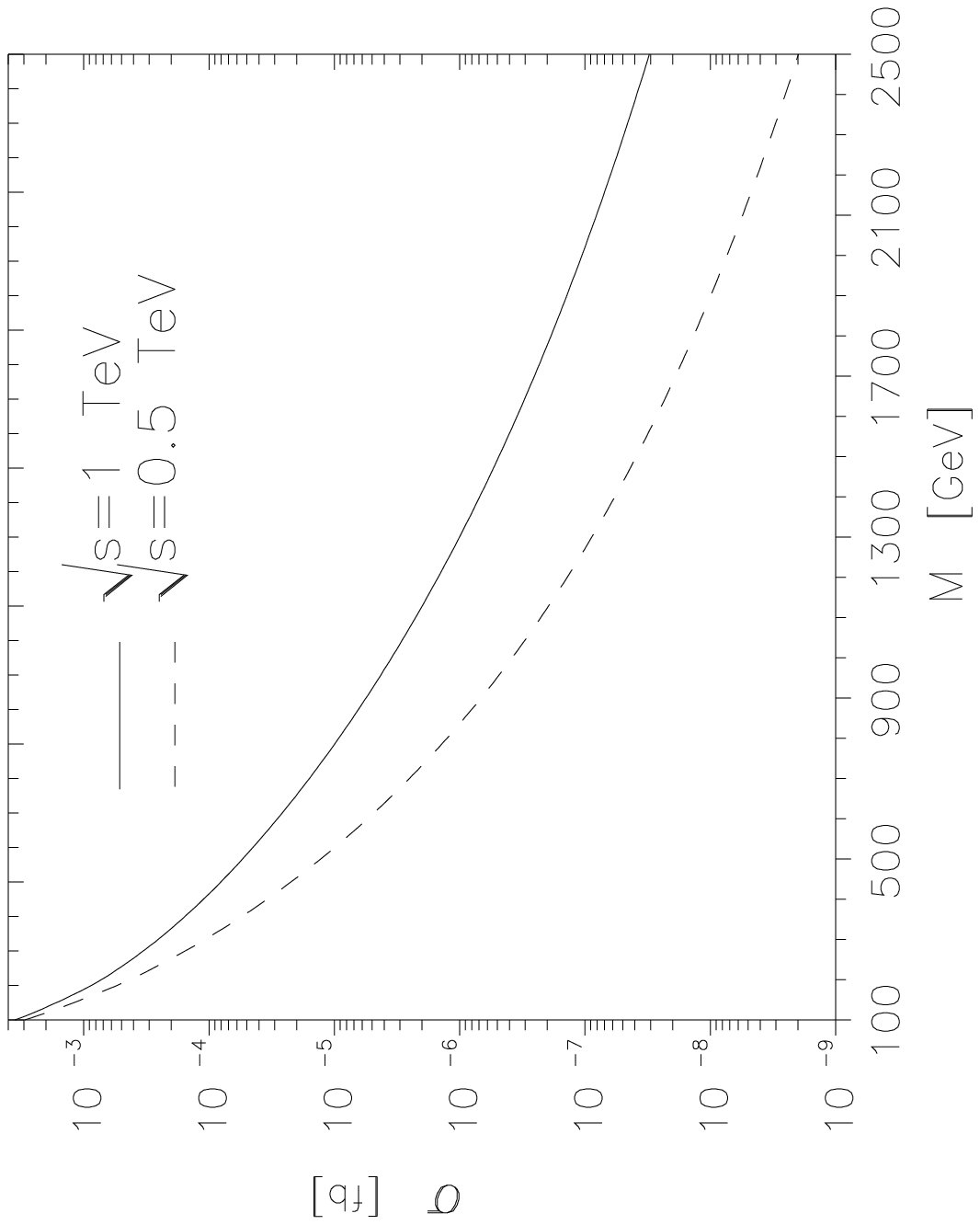}
\end{figure}
\baselineskip 5 mm
{\footnotesize Fig.15. The cross section for the $e^-e^- \rightarrow W^-W^-$ 
process as a function of the heavy neutrino mass for the classical `see-saw'
models, where the mixing angles between light and heavy neutrinos are
proportional to the inverse of mass of the heavy neutrino.} 

\baselineskip 7 mm

However, as we know from previous sections we can expect that results can be quite
different for `non-decoupling' models.

\subsection{`Non-decoupling model and the $e^-e^- \rightarrow W^-W^-$
process}

Now I would like to establish how big the utmost cross section for the $e^-e^- 
\rightarrow W^-W^-$ process can be. We can find in literature quite
optimistic estimations for that \cite{hmin}. I will show here
similarly to the $e^-e^+ \rightarrow \nu N$ case how `maximal' results
depend on the number of heavy neutrinos. 
\newpage
$\bullet \;n_R=1$ \\

If CP parities of all heavy neutrinos are the same or we have only one
right-handed neutrino then the $e^-e^- \rightarrow W^-W^-$
process is very small, much below $\sigma=$0.1 fb (here a very restrictive
bound on mixing angle $K_{Ne}$ (Eq.(2.23)) and the proportionality of the cross
section to the fourth power of this factor are crucial).

$\bullet \; n_R=2$ \\

We can have here large $K_{Ne}$ values
(Eq.(2.31)) but only when $A \rightarrow 1$ so two degenerate
Majorana neutrinos $(M_1=M_2)$ with opposite CP parities appear which
correspond to a one Dirac neutrino. 
As a Dirac neutrino conserves the lepton number the cross section vanishes.
It can be understood easily in a different way, too. Helicity amplitudes are
proportional to $K_{Ne}^2$ mixing (two the same vertices in t and u channels
(Fig.12)) and the amplitude is a sum over two exchanged heavy
neutrinos of equal mass. Then contributions from these two neutrinos are the same apart
from a relative sign which is opposite ($K_{Ne}$'s are pure real and complex
for them).

$\bullet \; n_R=3$ \\

The case with $n_R=3$ changes situation and the most optimistic results for
this case are shown in Fig.16. Taking $\eta_{CP}(N_1)=
\eta_{CP}(N_2)=-\eta_{CP}(N_3)=i$ and $M_1=M,\;M_2=AM,\;M_3=BM$ we can find
values A,B (masses of heavy neutrinos) for which mixings of heavy neutrinos are such 
(Eqs.(2.32)-(2.35))
that $\sigma( e^-e^- \rightarrow W^-W^-)
$ reaches the maximal value\footnote{Careful reader could noted contradiction between 
conclusions about CP effects
in previous subsection (`see-saw' model) and this given for the
`non-decoupling' model. In Figs.13,14 the solid line represents constructive
interferences among different heavy neutrinos contributions - cross section
is the biggest (CP parities of heavy neutrinos are the same). For the
`non-decoupling' model, however, biggest results can be obtained when
neutrinos have unlike CP parities (Section 2). The reason for this
is that for the `non-decoupling' model contribution of some heavy neutrino with
mixing angle $K_{Ne}$ dominates over others and CP effects as discussed for
the
`see-saw' model vanish ($K_{Ne}$'s in the `see-saw' model are comparable 
and interferences can
appear (Eq.(4.4)). As ${(K_{Ne})}_{`non-dec.'} >>
{(K_{Ne})}_{`see-saw'}$ and consequently $\sigma(e^-e^-\rightarrow
W^-W^-)_{`non-dec.'} >> \sigma(e^-e^-\rightarrow W^-W^-)_{`see-saw'}$ the
conclusions about CP effects are quite different.}. This situation takes place
for a very heavy second
$(A >>1)$ and a heavier third neutrino ($B \sim 2-10$). 
In this Figure I
depict also the cross section for production of the lightest heavy neutrinos
with the mass M in the $e^+e^- \rightarrow \nu N$ process taking exactly the same
mixing angle $K_{N_1e}$ as for the $e^-e^- \rightarrow W^-W^-$ process. \\

We can see that 

(i) everywhere in the possible region of phase space the
production of heavy neutrinos in the $e^+e^-$ process has 
greater cross section than the lepton violating process
$e^-e^-$. It is impossible to find such mixing angles and masses which 
would show the opposite.

(ii) there are regions of heavy neutrino masses outside
the phase space for their production in the $e^+e^-$ process
where the $\Delta L=2$ process
$e^-e^- \rightarrow W^-W^-$ is still a possible place to look for heavy neutrinos. It
is a small region $1\;{\rm TeV}<M<1.1\;{\rm TeV}$ for $\sqrt{s}=1$~TeV,
$1.5{\rm\;TeV}<M<2{\rm\;TeV}$ for $\sqrt{s}=1.5$ TeV and
$2{\rm\;TeV}<M<3.1{\rm\;TeV}$ for $\sqrt{s}=2$~TeV where the cross section
$\sigma \left( e^-e^- \rightarrow W^-W^- \right) $ is still above the `detection
limit'. There is no such place with the $\sqrt{s}=0.5$ TeV
collider. The experimental value of $\kappa^2$ (see Eq.(2.16)) would have to
be below $\sim 0.004,\sim 0.003,\sim 0.002$ for
$\sqrt{s}=1,1.5,2$ TeV respectively to cause these regions
to vanish. 

\newpage 

\ \\

\vspace{6 cm}
\begin{figure}[h]
\includegraphics{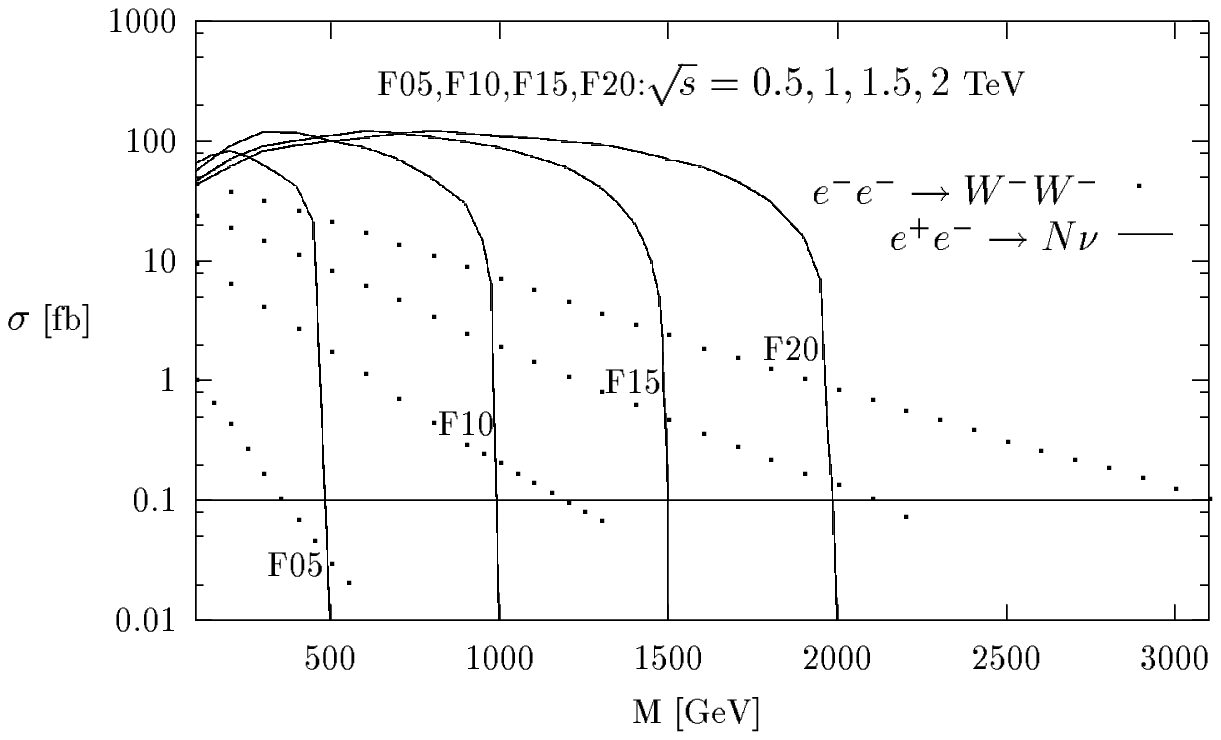}
\end{figure}

\baselineskip 5 mm
{\footnotesize Fig.16. The cross sections for the $e^+e^- \rightarrow N\nu$ and 
$e^-e^- \rightarrow W^-W^-$
processes as a function of the lightest neutrino mass $M_1=M$ for different CM
energies (the
curves denoted by F05, F10, F15 and F20 depicted the cross section for both
processes
for $\sqrt{s}=$0.5, 1, 1.5 and 2 TeV respectively) for $n_R=3$.
The cross sections for the $e^-e^- \rightarrow W^-W^-$ process are chosen to be
the largest. For the
$e^+e^- \rightarrow N\nu$ reaction the cross section for each of neutrino
masses is calculated using
the same parameters as for $\sigma (e^-e^- \rightarrow W^-W^-)$.
The solid line parallel to the
$M$  axis  gives the predicted `detection limit' $(\sigma=0.1\;$fb) for the
$e^-e^- \rightarrow W^-W^-$ process.} \\

\baselineskip 7 mm
 
The above results are exact for the RHS model. In the LR model the
situation is different as doubly charged Higgs particles exist and a
resonance can appear.
This situation is summarized on Fig.17 \cite{pr10}.

The total decay width for the $\delta_R^{--}$ is taken to be 10
GeV. We can see that even for CM energies $\sim 4 \Gamma_{\delta_R^{--}}$
out of resonance peak the s channel contribution dominates over t and u channel
ones. The reason is that for $\sqrt{s} \leq 500$ GeV contributions of t and u
channels are below $\sigma=0.1$ fb for almost all allowed space of
parameters (see Fig.16). Specially  promising results are when 
$M_{W_2}$ is very close to its experimental limit
(\cite{wr}). 

\newpage 

\ \\

\vspace{7.5 cm}
\begin{figure}[h]
\includegraphics{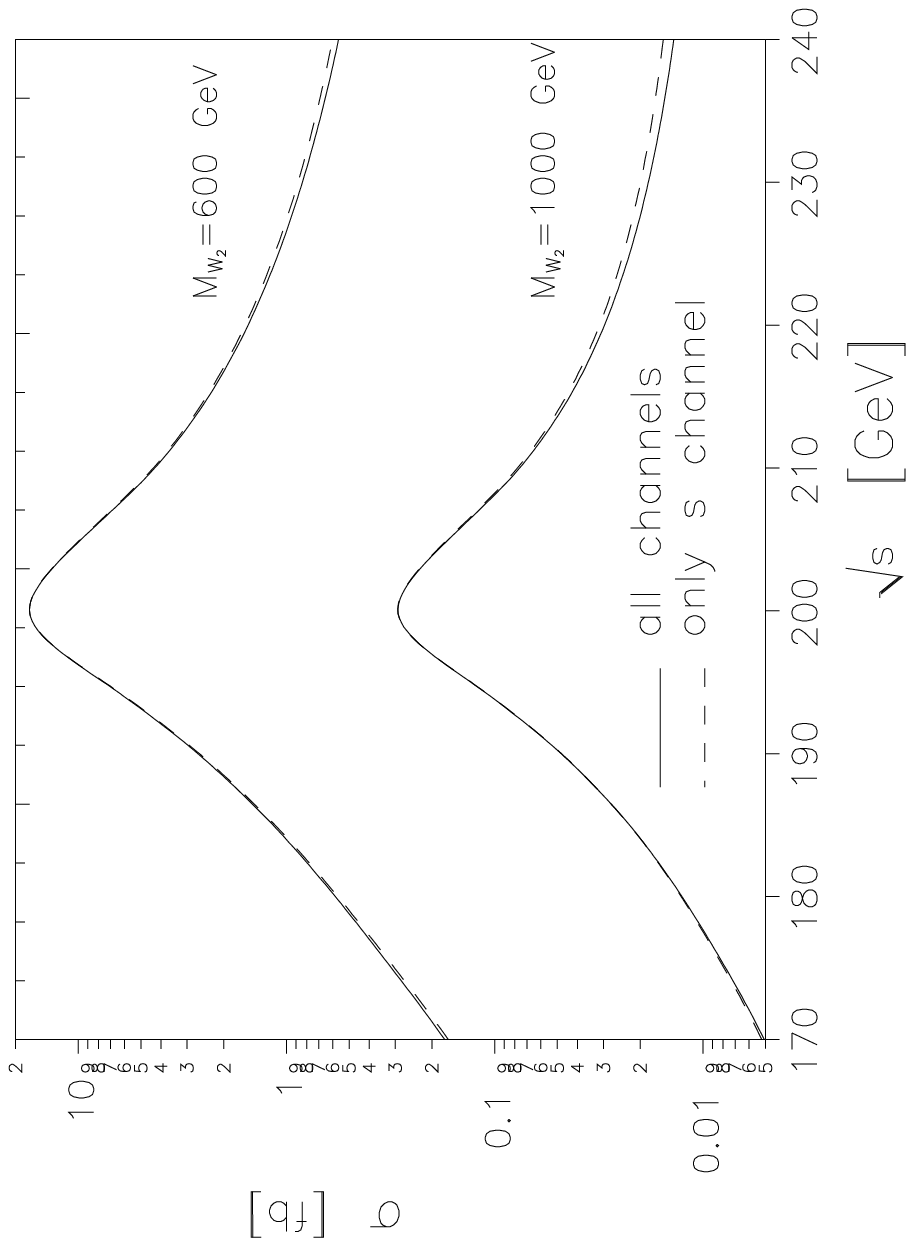}
\end{figure}

\baselineskip 5 mm

{\footnotesize Fig.17. 
The cross sections for the $e^-e^- \rightarrow W^-W^-$ process with
$\delta_R^{--}$ resonance ($M_{\delta_R^{--}}$=200 GeV) as a CM energy
function.
The solid line represents the total cross section with contribution from t,u
and s channels altogether, the dashed line depicts the cross section after
removing contributions of t and u channels.}

\baselineskip 7 mm


We would like to point out that doubly charged Higgs particle detection in s
channel
would also indicate that heavy neutrinos exist. 
The reason is that coupling of the $\delta_R^{--}$ Higgs particle with
electrons is proportional to neutrino masses and the light neutrinos alone 
are not sufficient to give detectable s-channel signal in the $e^-e^-
\rightarrow W^-W^-$ process \cite{pr10}.

\newpage

\setcounter{equation}{0}
\renewcommand{\theequation}{5.\arabic{equation}}
\section{Conclusions}

None of the nonstandard processes involving a heavy neutrino has ever been
detected. In this Thesis I have discussed possibilities of their detection by
the $e^-e^+ \rightarrow \nu N$ and the $e^-e^- \rightarrow W^-W^-$
processes which are the most promising 
reactions in the NLC. Cross sections for these processes
are very sensitive to the heavy neutrino mixing angle with an electron ($ \sigma (
e^-e^+ \rightarrow \nu N) \sim K_{Ne}^2$,  
$\sigma( e^-e^- \rightarrow W^-W^-)
\sim K_{Ne}^4 $). In the framework of the `see-saw' class of models this mixing angle is
connected with a heavy neutrino mass and because this mass is large 
$(> M_Z)$ the mixing angle is
small and decreases with a heavy neutrino mass. It causes that cross sections
are very small for the $e^-e^+ \rightarrow \nu N$ process and detectable
only for not too heavy neutrinos (Fig.6). CP effects are important when two
heavy neutrinos are produced (no helicity suppression factors above
threshold)  with the changes in the cross
sections which result from the CP breaking phases and which can be several 
times
larger than the cross sections themselves (Figs.4 and 5).
Although CP effects are quite interesting in the framework of the `see-saw'
model for the $e^-e^- \rightarrow W^-W^-$ process, too (Figs.13,14), the
values of cross sections are so small that there is practically no chance
for process detection (Fig.15).
However, there are other (`non-decoupling') models where
mixing angles are independent of heavy neutrino masses and can be large -
the only restriction on their values comes from experimental data.
The goal of this Thesis has been to show that taking into account all stringent
limits on heavy neutrino mixing angles and masses the heavy neutrinos can still
be detected through both the  
$e^-e^+ \rightarrow \nu N$ process (Figs.7-11) and the $e^-e^- \rightarrow W^-W^-$ 
process (Figs.16-17) in future linear colliders. 
From the results given there we can
also deduce how much those low-energy limits would have to change in future
to cause that these processes would be below detection and also, what
possibly detection would tell us about heavy neutrinos themselves. 
It appears that to get significant
cross sections the two characteristics of heavy neutrinos are crucial:
their number and CP eigenvalues. 
The detectable cross sections can be obtained only when $n_R>1$
and when heavy neutrinos have unlike CP parities. Then the cross
section for the $e^-e^+ \rightarrow \nu N$ process can be as large as
$\sigma \simeq 275$ fb for $n_R=2$ and $\sqrt{s}=1$ TeV 
(and similarly for $n_R>2$) and heavy neutrinos can be
discovered by electrons detection in the $e^-e^+ \rightarrow \nu (N
\rightarrow e^- W^+)$ chain (backward distribution).

The cross section for the $e^-e^- \rightarrow W^-W^-$ process is still under
detection when $n_R=2$. If $n_R=3$ then there is a possible space of heavy
neutrino-electron  mixing angles and heavy neutrino masses where the process 
$e^-e^- \rightarrow W^-W^-$
can be detected for $\sqrt{s} \geq 500$ GeV. For $\sqrt{s} \leq 500$ GeV 
the $\delta_R^{--}$ resonance (LR model) can enhance the
process ($e_R^-e_R^-$ collision) specially when 
a mass of the additional charged gauge boson is below 1 TeV.
The discovery of the doubly charged Higgs particle through the $e^-e^-$
s-channel resonance would be also an (indirect) indication that heavy neutrinos exist.

\newpage

\setcounter{equation}{0}
\renewcommand{\theequation}{A.\arabic{equation}}
\section{Appendix A. \
Neutrino mass matrices, their diagonalization and mixing matrices}

I consider two types of gauge models. 

\subsection*{RHS model}

This model is based on $SU(2)_L \otimes U(1)_Y$ gauge group. The left handed
lepton fields are put in doublets and transformed according to SU(2) 
representation (the quantum numbers of weak isospin and hypercharge 
$(T_{3L},Y)$ are given in brackets)
\begin{eqnarray}
L_{iL}= \left( \matrix{ \nu_{l_j} \cr l_j^- } \right)_L \matrix{: (1/2,-1) \cr
\;\;\;:(-1/2,-1) },\;\;\; l_j=e,\mu,\tau,
\end{eqnarray}
meanwhile the right-handed lepton fields are SU(2) singlets
\begin{eqnarray}
l_{jR}&=&e_R^-,\;\nu_R^-,\;\tau_R^-\;\;\;\;\;\;\;:(0,-2), \nonumber \\
\nu_{jR}&=&\nu_{1R},\; \nu_{2R},\;...\;\;\;\;\;\ :(0,0). 
\end{eqnarray}

The only difference between this model and the standard, GWS model is caused
by the presence of the right-handed neutrino fields. The remaining part of 
the model
is exactly the same (quark, Higgs sector). That means that the lepton part
of the Lagrangian allowed by the gauge
symmetry is the following

\begin{eqnarray}
L_{leptons}&=&i \bar{L}_{L} D L_{L}+i\bar{l}_{R} D l_{R}+
i\bar{\nu}_{R} D \nu_{R} \nonumber \\
&-&h^l [\bar{L}_{L} \tilde{\phi} l_{R} ] -
h^{\nu} [\bar{L}_{L} {\phi} \nu_{R} ] -\frac{1}{2}
\bar{\nu}_{L}^c{(M_R)}\nu_{R}+h.c.
\end{eqnarray}

where $\phi$ denotes a Higgs doublet $\left( \tilde{\phi}= \epsilon \phi^{\ast}=
\left( \phi^+,-\phi^{0\ast}  \right)^T \right)$

\begin{eqnarray}
\phi&=& \left( \matrix{ \phi^0 \cr \phi^- } \right) = \frac{1}{\sqrt{2}} 
\left( \matrix{ v+H^0 +i \chi^0 \cr \chi^- } \right) ,
\end{eqnarray}
 
$D=\gamma^{\mu}D_{\mu}$ and covariant derivative has the form
\begin{equation}
D_{\mu}=\left\{ 
\begin{array}{cl}  
\partial_{\mu}-ig \frac{\vec{\tau}}{2}\vec{W}_{\mu}
-\frac{ig'}{2}YB_{\mu} &  \;\; \mbox{for}\; \mbox{doublets} \\
\partial_{\mu}- \frac{ig'}{2}YB_{\mu} & \;\; \mbox{for} \; \mbox{singlets}.
\end{array}
\right.
\end{equation}
Matrices $h^l,h^{\nu}$ of $3 \times 3$ and $3 \times
(3+n_R)$ dimensions, respectively, describe couplings of lepton fields with
Higgs fields and 
the symbol $`v'$ in Eq.(A.4) describes Standard Model's vacuum expectation
value.
So, besides lepton-Higgs field couplings
the Lagrangian include mass terms for leptons
\begin{eqnarray}
L_{mass}&=&- \bar{l}_{L} m^l l_{R}-
\bar{\nu}_{L}m_D \nu_{R} -\frac{1}{2}\bar{\nu}_{L}^c{(M_R)} \nu_{R}+ h.c. \\
&&m_D=\frac{h^{\nu}v}{\sqrt{2}},\;\;\;\;\;\;\;
m^l=\frac{h^{l}v}{\sqrt{2}}.
\end{eqnarray}
Full mass Lagrangian for neutrinos can be rewritten as follows 
\begin{equation}
L_{mass}^{neutrinos}=-\frac{1}{2} \left( \bar{\nu}_{L}, \bar{\nu}_{L}^c \right)
\left( \matrix{ 0 & m_D \cr 
                m_D^T & M_R } \right) 
\left( \matrix{ \nu_{R}^c \cr \nu_{R} } \right) + h.c.
\end{equation}
where neutrino mass matrix $M_{\nu}$ has $(3+n_R) \times (3+n_R)$ dimension
\begin{equation}
M_{\nu}= { \overbrace{0}^{3} \ \overbrace{m_D}^{n_R} \choose
           m_D^T \ M_R }
      \matrix{ \}3 \cr \; \} n_R}.
\end{equation}                           
Perturbative calculations demand to have
$\mid h^{\nu} \mid_{ij}, \mid h^l \mid_{ij} \leq 1$, so all elements of the 
$m_D$ matrix ($3 \times 3$ dimension) are utmost of the order of charged 
lepton masses. 
Then, if we want to be consistent with the experimental data,
matrix $M_R$ 
of the $n_R \times n_R$ dimension must fullfil inequalities
$(m_D)_{ij} << (M_R)_{ij} \geq 100$ GeV. 

We can construct Majorana neutrinos ($\nu_M=\nu_M^C$)
\begin{equation}
\nu_M = \nu_R + \nu_L^c 
\end{equation}
and then, using unitary matrix U in the form
\begin{equation}
U= \overbrace{ \left( \matrix{ U_L^{\ast} \cr U_R } \right) }^{3+n_R} 
\matrix{ \} 3 \cr \; \} n_R} 
\end{equation}
we can diagonalize matrix $M_{\nu}$
\begin{equation}
U^TM_{\nu}U=M_{diag}.
\end{equation}
Thanks to this transformation we obtain simultaneously mixing matrix 
among weak ($\nu_M$) and physical ($N$) states
\begin{equation}
N_i=\sum\limits_{j=1}^{3+n_R} U_{ij} \nu_{M_{j}}.
\end{equation}
Eventually the mass Lagrangian takes a form
\begin{equation}
2L_{mass}^{neutrin}=-\sum_{i=1}^{3+n_R} \bar{N}_i (M_{diag})_{ii}N_i.
\end{equation}
Mass matrix for charged leptons $m^l$ is diagonalized by unitary
matrix $U_L^l$
\begin{eqnarray}
L^{leptons}&=&-\sum\limits_{l=1}^3 m_{diag}^l\bar{\hat{l}}\hat{l}, \\
{\rm where} \hspace{2 cm} && \nonumber \\
\hat{l}&=&U_L^ll.
\end{eqnarray}
\subsection*{LR model} 

This model is based on the $SU(2)_L \otimes SU(2)_R \otimes
U(1)_{B-L}$ gauge group. Lepton sector is the reflection of this symmetry.
The left and right lepton (and quark) states are included in doublets
(the quantum numbers connected with $SU(2)_L$, $SU(2)_R$ and $U(1)$ 
gauge groups, ($T_{3L},T_{3R},Y$) respectively, are included in brackets)
\begin{equation}
\Psi_{iL}=\left( \matrix{ \nu_i \cr l^-_i } \right)_L \;\;:\;(2,1,-1),\; 
\Psi_{iR}=\left( \matrix{ \nu_i \cr l^-_i } \right)_R :\;\;\;(1,2,-1)
\end{equation}
where index i for neutrinos and $l_i$ for leptons stand for 
$e,\mu,\tau$.

Higgs sector can be realized on many ways in this model \cite{sechig}. 
Classical model \cite{tripsec} which I use here includes a bidoublet
and two triplets

\begin{eqnarray}
\phi&=& \left( \matrix{ \phi_1^0 & \phi_1^+ \cr
                      \phi_2^- & \phi_2^0 \cr} \right), \\
&& \nonumber \\                                            
\Delta_{L,R}&=& \left( \matrix{ \delta_{L,R}^+/\sqrt{2} & \delta_{L,R}^{++} \cr
                          \delta_{L,R}^0 & -\delta_{L,R}^+/\sqrt{2} \cr }
                          \right) . 
\end{eqnarray}

Spontaneous symmetry breaking mechanism minimalizes Higgs potential for the 
following choice of the vacuum expectation values of the  
$\delta_{R,L}^0\;,\;\phi_{1,2}^0$ fields \cite{tripsec}
\begin{equation}
<\Phi>=\left( \matrix{ \kappa_1/\sqrt{2} & 0 \cr
                    0 & \kappa_2/\sqrt{2} } \right) \;\;\;,\;\;\;
<\Delta_{L,R} > = \left( \matrix{ 0 & 0 \cr
                      v_{L,R}/\sqrt{2} & 0 } \right).
\end{equation}          

In this model
the lepton part of the Lagrangian allowed by gauge symmetry takes the form
\begin{eqnarray}
L_{leptons}&\equiv&L_Y^B+L_Y^L+L_Y^R=i\bar{\Psi}_LD\Psi_L+i\bar{\Psi}_RD\Psi_R
-\bar{\Psi}_L \left[ h\phi + \tilde{h}
\tilde{\phi} \right] \Psi_R \nonumber \\
&-& \bar{\Psi}_LCi\tau_2 h_L\Delta_L\Psi_L 
- \bar{\Psi}_RCi\tau_2h_R\Delta_R\Psi_R 
\end{eqnarray}
where
$$\tilde{\phi}=\tau_2 \phi^{\ast} \tau_2,$$
$D=\gamma^{\mu}D_{\mu}$  and $D_{\mu}$ equals
\begin{equation}
D_{\mu}=\left\{ 
\begin{array}{cl}  
\partial_{\mu}-ig \frac{\vec{\tau}}{2}\vec{W_L}_{\mu}
+\frac{ig'}{2}YB_{\mu} &  \;\; \mbox{for}\;\mbox{left}\;\mbox{spinors} \\
\partial_{\mu}-ig \frac{\vec{\tau}}{2}\vec{W_R}_{\mu}
+\frac{ig'}{2}YB_{\mu} &  \;\; \mbox{for}\;\mbox{right}\;\mbox{spinors}.
\end{array}
\right.
\end{equation}

The left-right symmetry forces the following relations
\begin{equation}
h=h^{\dagger}\;\;\;,\;\;\;\tilde{h}=\tilde{h}^{\dagger}\;\;\;,\;\;\;h_L=h_R.
\end{equation}

Let's assume that left triplet does not condensate 
$(v_L=0)$ - this ensures that unnaturally small values of parameters in Higgs
potential are absent. Discussion of necessity (and naturality) for such
choice has been carried out in \cite{pr3}, \cite{tripsec}\footnote{ The interest 
is in $\beta_1,\beta_2,\beta_3$ parameters
which multiply terms which mix $\Delta_L$ fields with $\Delta_R$ ones, 
for instance (\cite{tripsec}, (A.2)): 
$\beta_1 (Tr[ \phi \Delta_R 
\phi^{\dagger} \Delta_L^{\dagger}]+Tr[ \phi^{\dagger} \Delta_L 
\phi \Delta_R^{\dagger}])$. Consistency between potential minimalization
(which gives relation where $v_L \sim \beta_i$) and any allowed spectrum
of light neutrino masses (mass matrix for light neutrinos is on the other
hand proportional to $v_L$) demand $\beta_i \leq 10^{-6}$
\cite{tripsec}. Discrete symmetry of the type $\Delta_L \rightarrow
\Delta_R, \;\Delta_R \rightarrow - \Delta_L$ can eliminate this difficulty
\cite{pr3}.}
and then the mass matrix for neutrinos is of the type
\begin{equation}
M_{\nu}= { \overbrace{0}^{3} \ \overbrace{m_D}^{3} \choose
           m_D^T \ M_R }
      \matrix{ \}3 \cr \; \} 3}.
\end{equation}                           
where
\begin{equation}
m_D=\frac{1}{\sqrt{2}}\left( h\kappa_1+\tilde{h}\kappa_2 \right)\;\; \mbox{\rm and}\;\;
M_R=\sqrt{2}h_Rv_R
\end{equation}
are hermitian and symmetric matrices of dimension 3, respectively.
Masses of charged leptons come from the same Yukawa couplings
\begin{equation}
m^l=\frac{1}{\sqrt{2}}\left( h\kappa_2+\tilde{h}\kappa_1 \right).
\end{equation}
Similarly to the previous model we can define matrices $U,U^l_L$ which
transform weak neutrino and charged lepton states to the physical ones and
to get mass matrices as in Eq.(A.14) (with $n_R=3$) and in Eq.(A.15).

Together we get three light neutrinos and three heavy ones of the order
of $v_R\;\;(v_R>>\kappa_1,\kappa_2)$.
\setcounter{equation}{0}
\renewcommand{\theequation}{B.\arabic{equation}}
\section{Appendix B. \
Couplings of neutrinos in charged and neutral currents. Couplings of neutrinos
with Higgs particles}

In this Appendix I describe all relevant couplings which must be known
for numerical calculations of the considered processes $e^-e^+ \rightarrow
N_a N_b$ (Section 3) and $e^-e^- \rightarrow W^-W^-$ (Section 4).

\subsection*{RHS model} 

Charged and neutral currents have the form as in the SM
\begin{eqnarray}
L_{CC}&=&\frac{g}{\sqrt{2}} \bar{\nu}_{iL}\gamma^{\mu}l_{iL}W_{\mu}^+ 
+ h.c., \\
L_{NC}&=&\frac{g}{2\cos{\theta_W}} Z_{\mu}
\bar{\nu}_{iL}\gamma^{\mu}\nu_{iL}.
\end{eqnarray}
These interactions can be written using relations (A.13),(A.16) in a base
of physical states
\begin{eqnarray}
L_{CC}&=&\frac{g}{\sqrt{2}}\bar{N}\gamma^{\mu}KP_L\hat{l}W_{\mu}^+ + h.c., \\
L_{NC}&=& 
\frac{g}{2\cos{\theta_W}} Z_{\mu} \sum_{a,b} \bar{N}_a  \gamma^{\mu}P_L
\Omega_{ab} N_b \nonumber \\
&=&\frac{g}{2\cos{\theta_W}} Z_{\mu}  \frac{1}{\delta_{ab}+1}
\sum_{a \geq b} \bar{N}_a 
\left[ \gamma^{\mu}(P_L\Omega-P_R\Omega^{\ast}) \right]_{ab} N_b 
\end{eqnarray}
where I have used the following denotations\footnote{without loosing   
generality the matrix $U_L^l$ can be chosen as the identity one \cite{bil} 
and such approach is used in this Thesis.}
\begin{equation}
K=U_L^TU_L^l,\;\;\;\;\Omega=U_L^{\dagger}U_L.
\end{equation}
Similarly, writing down Yukawa interaction in physical states we get
\begin{eqnarray}
L_{Yukawa}^{D-D}&=&-
\frac{g}{2M_W} \sum_a \{ m_a^l (\bar{l}_a l_aH^0+i\bar{l}_a\gamma_5 l_a\chi^0)\} 
\nonumber \\
L_{Yukawa}^{D-M}&=&-\frac{g}{2M_W} \chi^- \left( \sum_{a,b} \bar{l}_a [ P_R 
(K^{\dagger})_{ab}m_b^N-P_Lm_a^l(K^{\dagger})_{ab} ] N_b \right) + h.c.
\nonumber
\end{eqnarray}
\begin{equation}
\end{equation}
for vertices which include two Dirac fermion particles (`Dirac-Dirac' vertex) or
one Dirac and one Majorana particles (`Dirac-Majorana' vertex).

For `Majorana-Majorana' vertices:
\begin{eqnarray*}
L_{Yukawa}^{M-M}(H^0)&=&-\frac{g}{2M_W}H^0
\sum_{a,b} \left( \bar{N}_a \Omega_{ab}m_b^NP_R N_b+\bar{N}_a 
m_a^N\Omega_{ab}P_L N_b \right)  \\
&=&-\frac{g}{2M_W}H^0 \frac{1}{\delta_{ab}+1}  \\
&&\sum_{a \geq b}
\bar{N}_a \left[ P_R(\Omega_{ab}m_b^N+\Omega_{ab}^{\ast}m_a^N)+
P_L(\Omega_{ab}m_a^N+\Omega_{ab}^{\ast}m_b^N) \right] N_b  
\end{eqnarray*}
\begin{equation}
\end{equation}
\begin{eqnarray*}
L_{Yukawa}^{M-M}(\chi^0)&=&
-\frac{ig}{2M_W}\chi^0 
\sum_{a,b} \left( \bar{N}_a \Omega_{ab}m_b^NP_R N_b-\bar{N}_a 
m_a^N\Omega_{ab}P_L N_b \right) \\
&=&-\frac{ig}{2M_W}\chi^0 \frac{1}{\delta_{ab}+1} \\
&& \sum_{a \geq b}
\bar{N}_a \left[ P_R(\Omega_{ab}m_b^N+\Omega_{ab}^{\ast}m_a^N)-
P_L(\Omega_{ab}m_a^N+\Omega_{ab}^{\ast}m_b^N) \right] N_b  
\end{eqnarray*}
\begin{equation}
\end{equation}
\subsection*{LR model} 

This model involves two pairs of charged and neutral gauge bosons.
Their masses come from the kinetic part of the Higgs Lagrangian
\begin{equation}
L_{kinet}=Tr{(D^{\mu}\phi)}^{\dagger}(D_{\mu}\phi)+
Tr{(D^{\mu}\Delta_L)}^{\dagger}(D_{\mu}\Delta_L)+
Tr{(D^{\mu}\Delta_R)}^{\dagger}(D_{\mu}\Delta_R)
\end{equation}
where 
\begin{eqnarray}
D^{\mu}\phi&=&\partial_{\mu}\phi-ig\frac{\vec{\tau}}{2}\phi \vec{W}_{L\mu}
+ig\phi\frac{\vec{\tau}}{2}\vec{W}_{R\mu}   \nonumber \\
D^{\mu}\Delta_{L,R}&=& \partial_{\mu}\Delta_{L,R}-ig\vec{W}_{L,R\mu}
[\frac{\vec{\tau}}{2},\Delta_{L,R}]-ig'B_{\mu}\Delta_{L,R}
\end{eqnarray}
Writing at length this Lagrangian we obtain after spontaneous symmetry
breaking  (Eq.(A.20))
mass matrices for gauge
fields $(\kappa_+^2=\kappa_1^2+\kappa_2^2)$
\begin{eqnarray}
\tilde{M}_W^2&=& \frac{g^2}{4} \left( \matrix{ \kappa_+^2 & -4\kappa_1\kappa_2
 \cr
                      -4\kappa_1\kappa_2 & \kappa_+^2+2v_R^2 } \right), \\
                     &&  \nonumber \\
\tilde{M}_0^2&=& \left( \matrix{ \frac{g^2}{2}\kappa_+^2 & - \frac{g^2}{2}
                                  \kappa_+^2 & 0 \cr
            -\frac{g^2}{2}\kappa_+^2 & \frac{g^2}{2}(\kappa_+^2+4v_R^2) 
            & -4gg'v_R^2 \cr
            0 & -4gg'v_R^2 & 4g'^2v_R^2 } \right).
\end{eqnarray}            
Let's take unitary matrices which transform weak gauge boson states
to physical states in the form
\begin{equation}
\left(\matrix{W_{1}^\pm \cr
           \ \ \        \cr
       W_{2}^\pm \cr}\right)=
\left(\matrix{\cos {\zeta} \ &\ - \sin {\zeta} \cr
            \ \ \                \cr
        \sin {\zeta} \ &\  \cos {\zeta} \cr}\right)\
\left(\matrix{W_{L}^\pm \cr
     \ \ \           \cr
     W_{R}^\pm \cr}\right)\ ,
\end{equation}                 
for the charged sector and
\begin{equation}
\left(\matrix{Z_{1}\cr Z_{2}\cr A\cr}\right)=
\left(\matrix{c_{W}c,& -s_{W}c_{M}c-s_{M}s,&-s_Ws_{M}c+c_Ms  \cr
             c_{W}s,&-s_{W}c_{M}s+s_{M}c,& -s_Ws_Ms-c_Mc  \cr
         s_W, & c_Wc_M, & c_Ws_M  \cr}\right)
\left(\matrix{W_{3L}\cr W_{3R}\cr B }\right),\       
\end{equation}       
for the neutral one, where
\begin{eqnarray*}
c_W &=& \cos {{\Theta}_W },\;\;s_W=\sin {{\Theta}_W },\;\;c=\cos {\phi},\;\;s=\sin {\phi}, \\
c_M&=&\tan {{\Theta}_W },\;\;s_M= \frac{\sqrt{\cos{2{\Theta}_W }}}{\cos{{\Theta}_W}}.
\end{eqnarray*}
\ \ \ Transformations (B.13) and (B.14) define two mixing angles 
$\zeta, \phi$
and masses of gauge bosons which with good approximation are 
(\cite{pr3},\cite{pr2})
\begin{eqnarray}
\zeta&\simeq& \frac{2\kappa_1\kappa_2}{\kappa_1^2 +\kappa_2^2}\frac{M_{W_1}^2}{M_{W_2}^2},\; 
\phi\simeq-\frac{(\cos{2\Theta_W})^{3/2}}{2\cos^4{\Theta_W}}\frac{M_{W_1}^2}
{M_{W_2}^2} \nonumber \\
M_{W_{1}}^2&\simeq&\frac{g^2}{2}(\kappa_1^2+\kappa_2^2),\;M_{W_{2}}^2\simeq
\frac{g^2}{2}v_R^2, \\
M_{Z_{1}}^2&\simeq&\frac{g^2}{2\cos^2{\Theta_W}}(\kappa_1^2+\kappa_2^2),\;
M_{Z_{2}}^2\simeq\frac{2g^2\cos^2{\Theta_W}}{\cos{2\Theta_W}}v_R^2.
\end{eqnarray}
Let's note that relations (A.25) with $h_R \leq 1$ and (B.15) gives upper
bound on heavy neutrino masses
\begin{equation}
M_N \leq \frac{2M_{W_2}}{g}.
\end{equation}
Using transformations (A.13), (A.16)  we can state charged and neutral currents
as follows \cite{pr2}
\begin{equation}
L_{CC}=\frac{g}{\sqrt{2}} \sum_{i=1}^2 \bar{N} {\gamma}^{\mu}
\left[ A_L^{(i)}P_L+A_R^{(i)}P_R\right]\hat{l}W^+_{i\mu} +h.c.,
\end{equation}
and
\begin{eqnarray}
L_{NC}^{D-D}&=&\frac{g}{2\cos{\Theta_W}} \sum_{i=1}^2 \bar{\hat{l}}
\gamma^{\mu}[A^{il}_LP_L+A^{il}_RP_R ] \hat{l} Z_{i\mu}
\end{eqnarray}
for `Dirac-Dirac' vertex. For `Majorana-Majorana' ones we have
\begin{eqnarray*}
L_{NC}^{M-M}&=&
\frac{g}{2\cos{\Theta_W}} Z_{i\mu}\sum_{a,b} \bar{N}_a \gamma^{\mu} (P_L 
(\Omega_L)_{ab}A_L^{i\nu}+P_R(\Omega_R)_{ab}A_R^{i\nu})N_b \\
&=&\frac{g}{2\cos{\Theta_W}} Z_{i\mu}\frac{1}{\delta_{ab}+1} \\
& \times & \sum_{a \geq b} \bar{N}_a \gamma^{\mu} 
[(A_L^{i\nu}+A_R^{i\nu})(P_L
(\Omega_L)_{ab}-P_R(\Omega_L)_{ab}^{\ast})+(P_R-P_L)A_R^{i\nu}\delta_{ab}]N_b 
\end{eqnarray*}
\begin{equation}
\end{equation}
where (a stands for leptons (a=e,$\mu,\tau$) and $\beta$ stands for
neutrinos ($\beta=1,...,6$))
\begin{equation}
\begin{array}{lcllcl}
\left(A_L^{(1)} \right)_{a\beta}&=&\cos{\xi} \left(K_L\right)_{a\beta},\;&
\left(A_R^{(1)}\right)_{a\beta}&=&- \sin{\xi}\left(K_R\right)_{a\beta} ,\\
\left(A_L^{(2)}\right)_{a\beta}&=& \sin{\xi} \left(K_L\right)_{a\beta}
,&\left(A_R^{(2)}\right)_{a\beta}&=& \cos{\xi} \left(K_R\right)_{a\beta} , 
\end{array}
\end{equation}
and 
$$A_{L(R)}^{il(\nu)}=g_{L(R)}^{il(\nu)}
\cos{\phi}+{g'}_{L(R)}^{il(\nu)}\sin{\phi},$$
\begin{eqnarray}
g_L^{1l}&=&{g'}^{2l}_L=-1+2\sin^2{\Theta_W} \\
{g'}_L^{1l}&=&-g^{2l}_L={g'}^{1\nu}_L=-g^{2\nu}_L=-\frac{\sin^2{\Theta_W}}
{\sqrt{\cos{2\Theta_W}}} \\
g_L^{1\nu}&=&{g'}^{2\nu}_L=1\;\;g_R^{1\nu}={g'}^{2\nu}_R=0 \\
g_R^{1l}&=&{g'}^{2l}_R=2\sin^2{\Theta_W} \\
{g'}_R^{1l}&=&-g{2l}_R=-\frac{1-3\sin^2{\Theta_W}}
{\sqrt{\cos{2\Theta_W}}} \\
{g'}_R^{1\nu}&=&-g{2\nu}_R=-\frac{\cos^2{\Theta_W}}
{\sqrt{\cos{2\Theta_W}}} 
\end{eqnarray}

Let's note that the neutral part of this Lagrangian for charged leptons 
conserves the lepton flavour.
Mixing matrices $K_{L,R}$ and  $\Omega_{L,R}$ have $6\times3$, $6 \times 6$ 
dimensions, respectively and are connected with transformation of fields
to physical states\footnote{Similarly to the RHS model (Eq.(B.5)) we can
choose charged lepton mixing matrices $U_{L,R}^l$ as identity. We denote
matrices $K_L$ and $\Omega_L$ which enter in left-handed currents exactly
as in the RHS model to make discussion of their meaning common through
this Thesis (Sections 3 and 4).}
\begin{equation}
K_L \equiv K =U_L^{\dag}U_L^l\ ,\ \ K_R=U_R^{\dag}U_R^l, \ \ 
\Omega_L \equiv \Omega =U_L^{\dagger}U_L,
\ \ \Omega_R=U_R^{\dagger}U_R.
\end{equation}

In the paper I consider the influence of the LR model's Higgs sector on 
the $e^+e^- \rightarrow \nu N$ cross section.
The value of this cross section depends both on the Higgs bosons masses and
on their couplings with
leptons. The form of the Higgs potential from which Higgs particles' masses 
can be obtained is large. Full discussion of the Higgs potential 
diagonalization and their couplings with neutrinos have been performed in
\cite{pr3} and \cite{tripsec}. I will restrict here only to showing indispensable
relations for 
$e^+e^- \rightarrow \nu N$ analysis.

Higgs fields give altogether 20 degrees of freedom. 
Minimalization of the Higgs potential gives 16 different particles
(denotations according to \cite{pr3}):
\begin{itemize}
\item 4 Goldstone' bosons ($G_{L,R}^{\pm},G_{L,R}^0$),
\item 6 neutral Higgs bosons ($H_0^0,H_1^0,H_2^0,H_3^0,A_1^0,A_2^0$),
\item 2 single charged particles ($H_1^{\pm},H_2^{\pm}$),
\item 2 doubly charged particles ($\delta_{L,R}^{\pm\pm}$).
\end{itemize}
Considering the existing experimental data it is justified to assume that
\begin{equation}
v_R >> y=\sqrt{\kappa_1^2+\kappa_2^2}
\end{equation}
and, taking into account
energies for which we make investigations, $\sqrt{s} >>m_e \simeq 0$. 
Then it turns out that only two neutral Higgs particles $H_1^0,A_1^0$ 
and two charged ones $H_1^-,H_2^-$ have nonzero couplings with leptons
(Fig.2). Masses of these particles are
\begin{eqnarray}
M_{H_1^{\pm}}^2&=&\frac{1}{2} \left[ v_R^2 + \frac{1}{2}y^2\sqrt{1-\epsilon^2}
\right], \nonumber \\
M_{H_2^{\pm}}^2&=&\frac{1}{2} \left[ v_R^2\frac{1}{\sqrt{1-\epsilon^2}} + 
\frac{1}{2}y^2\sqrt{1-\epsilon^2} \right], \nonumber \\
M_{H_1^{0}}^2&\simeq&\frac{1}{2}  v_R^2\frac{1}{\sqrt{1-\epsilon^2}}, \nonumber \\
M_{A_1^{0}}^2&\simeq&\frac{1}{2} \left[ v_R^2\frac{1}{\sqrt{1-\epsilon^2}} 
-4y^2 \right] \\
\mbox{\rm where} \hspace{1 cm} && \nonumber \\
0 \leq & \epsilon &
=\frac{2\kappa_1\kappa_2}{\kappa_1^2+\kappa_2^2} \leq 1. \nonumber
\end{eqnarray}

Their 
couplings with leptons (Eq.(A.21) are (symbols $\Gamma^{(x)}$, x=lN,l,N are
taken from Appendix C)

\begin{itemize}
\item $H_1^{\pm}$' exchange ($X \equiv
K^{\dagger}K_R^TM_{diag}^{\nu}K_RK^{\ast}$)
\end{itemize}
\begin{eqnarray}
\Gamma_{lN}\left( H_1^{\pm} \right)_{N_be^-}&=&\frac{1}{v_R}\sum_{c=4,5,6}
X_{bc}\left( K \right)_{ce}P_L, \nonumber \\
\Gamma_{lN}\left( H_1^{\pm} \right)_{e^+N_a}&=&\frac{1}{v_R}\sum_{c=4,5,6}
\left( K^{\dagger} \right)_{ec}\left( X^{\ast} \right)_{ca}P_R.
\end{eqnarray}
\begin{itemize}
\item $H_2^{\pm}$' exchange $\left( \alpha_2=\frac{\sqrt{2}}{y\sqrt{1-
\epsilon^2}} \right)$.
\end{itemize}
\begin{eqnarray}
&&\Gamma_{lN}\left( H_2^{\pm} \right)_{N_be^-}=
-\left[ m_b^N\left( K\right)_{be}\epsilon\alpha_2 \right]P_L + \\
&&\left[ -\sum_{c=4,5,6} \left(
\Omega_L \right)_{bc}m_c^N \left(K_R\right)_{ce}\alpha_2 
+ \frac{1}{\alpha_2v_R^2}\sum_{c=4,5,6} \left(
\Omega_R^{\ast} \right)_{bc}m_c^N \left(K_R\right)_{ce} \right] P_R , \nonumber \\
&& \nonumber \\
&&\Gamma_{lN}\left( H_2^{\pm} \right)_{e^+N_a}=
-\left[ \left(K^{\dagger} \right)_{ea}m_a^N\epsilon\alpha_2 \right]P_R + \\
&& \left[ -\sum_{c=4,5,6} \left(
K_R^{\dagger}\right)_{ec} m_c^N \left( \Omega \right)_{ca}\alpha_2 
+ \frac{1}{\alpha_2v_R^2}\sum_{c=4,5,6} \left(
K_R^{\dagger} \right)_{ec} m_c^N \left( \Omega_R^{\ast} \right)_{ca} \right] P_L. 
\nonumber
\end{eqnarray}
\begin{itemize}
\item neutral particles' exchange $H_0^0,H_1^0$ and $A_1^0$ 
\end{itemize}

Let's denote
\begin{eqnarray}
A_0&\simeq&\frac{1}{y\sqrt{1-\epsilon^2}}\left[H_1^0-iA_1^0\right], \nonumber \\
B_0&\simeq& \frac{1}{y}H_0^0-\frac{\epsilon}{y\sqrt{1-\epsilon^2}} \left[
H_1^0+iA_1^0\right],
\end{eqnarray}
then couplings $\left(e^-e^+H\right)$ and $\left(N_aN_bH\right)$
can be written in the form
\begin{eqnarray}
\Gamma_l \left( {\bf \{H_0^0\}},H_1^0,A_1^0\right)_{e^-e^+}
&=&-\left[ \sum_{c=4,5,6}
\left( K^{\dagger} \right)_{ce} \left( K_R \right)_{ce} m_c^N A_0 \right] P_R \nonumber \\
&&-\left[ \sum_{c=4,5,6}
\left( K^{\ast}\right)_{ce} \left( K_R^{\ast} \right)_{ce} m_c^N A_0^{\ast} \right] P_L,
\end{eqnarray}
(let's note that the lightest Higgs particle $H_0^0$ does not couple with
$e^-e^+$ pair if we neglect the mass of electron)
\begin{eqnarray}
\left[ \Gamma_N + \Gamma_N^C \right]_{N_aN_b}\left( H_0^0,H_1^0,A_1^0 \right)
&=& -[ ( \left( \Omega \right)_{ba}m_a^N+\left(\Omega
\right)_{ab} m_b^N ) B_0  \nonumber \\
&+&  \sum_{l=\mu,\tau}m_l \left( \left( K \right)_{bl}
\left(K_R^{\ast} \right)_{al}+\left( K \right)_{al}
\left(K_R^{\ast} \right)_{bl} \right)A_0^{\ast} ]P_R \nonumber \\
&-&[ ( \left( \Omega \right)_{ba}^{\ast}m_a^N+\left(\Omega
\right)_{ba} m_b^N ) B_0^{\ast} \nonumber \\
&+&  \sum_{l=\mu,\tau}m_l\left( \left( K^{\ast}
\right)_{bl} \left(K_R \right)_{al}+\left( K^{\ast} \right)_{al}
\left(K_R \right)_{bl} \right)A_0 ]P_L. \nonumber
\end{eqnarray}
\begin{equation}
\end{equation}
\newpage
\setcounter{equation}{0}
\renewcommand{\theequation}{C.\arabic{equation}}
\section{Appendix C. \
Feynman rules for Majorana neutrino interactions}
Majorana neutrinos are self-conjugate spin one half fields. Their self-conjugacy 
is responsible for the existence of four different propagators
in contrary to the Dirac neutrino case where only one propagator exists. This property
gives also bigger number of vertices and the problem with establishing relative 
signatures of various diagrams contributing to a given amplitude. We can however,
similarly to the Dirac case, introduce the Feynman diagram technique
which in a consistent way describes Majorana particles both for charged
and neutral currents.

Let's define a general form of Majorana particle interactions with gauge fields
in the following way (N - Majorana neutrino, l - charged lepton,
$W^{\pm}$,$Z^0$ - gauge bosons)
\begin{eqnarray}
L_{CC}&=&\bar{N}\Gamma^{\mu}_llW_{\mu}^++\bar{l} \bar{\Gamma}^{\mu}_lNW_{\mu}^-
\\
L_{NC}&=&\bar{l}\Gamma^{\mu}_{lN}lZ_{\mu}^0+\bar{N} \bar{\Gamma}^{\mu}_NNZ_{\mu}^0
=\bar{l}\Gamma^{\mu}_{lN}lZ_{\mu}^0+Z_{\mu}^0\sum_{a,b}\bar{N}_a 
\left( \Gamma^{\mu}_N \right)_{ab} N_b \nonumber \\
&=&\bar{l}\Gamma^{\mu}_{lN}lZ_{\mu}^0+Z_{\mu}^0 \frac{1}{\delta_{ab}+1}
\left( \sum_{a \geq b}\bar{N}_a {\left( \Gamma^{\mu}_N \right)}_{ab} N_b+ 
\bar{N}_b {\left( \Gamma^{\mu}_N \right)}_{ba} N_a  \right) \nonumber \\
&=& \bar{l}\Gamma^{\mu}_{lN}lZ_{\mu}^0+Z_{\mu}^0 \frac{1}{\delta_{ab}+1}\left(
\sum_{a \geq b}\bar{N}_a \left[ { \Gamma^{\mu}_N }
+{ \Gamma^{\mu}_N }^C \right]_{ab} N_b \right)
\end{eqnarray}
where
\begin{eqnarray}
\Gamma^{\mu}_{(x)}&=&\gamma^{\mu} \left( P_LA_L^{(x)}+P_RA_R^{(x)} \right),\;
x=l,N,lN, \\
\bar{\Gamma}^{\mu}_l&=&\gamma_0  {\Gamma^{\mu}_l}^{\dagger}\gamma_0 \\
{\rm and} \hspace{2 cm} && \nonumber \\
{\Gamma^{\mu}_N} ^C&=&C {\Gamma^{\mu}_N}^T C^{-1}.
\end{eqnarray}
Making algebraic manipulations in Eq.(C.2) we have used self-conjugacy of
Majorana field $(N^C \equiv C\bar{N}^T=N)$.
Similarly, let's define Majorana particle - Higgs boson $(H^{\pm},H^0)$ 
interaction
\begin{eqnarray}
L_{NH^{\pm}}&=&\bar{N}\Gamma_llH^++\bar{l} \bar{\Gamma}_lNH^-
\\
L_{NH^0}&=&\bar{l}\Gamma_{lN}lH^0+\bar{N} \bar{\Gamma}_NNH^0
=\bar{l}\Gamma_{lN}lH^0+H^0\sum_{a,b}\bar{N}_a 
\left( \Gamma_N \right)_{ab} N_b \nonumber \\
&=&\bar{l}\Gamma_{lN}lH^0+H^0 \frac{1}{\delta_{ab}+1}
\left( \sum_{a \geq b}\bar{N}_a {\left( \Gamma_N \right)}_{ab} N_b+ 
\bar{N}_b {\left( \Gamma_N \right)}_{ba} N_a  \right) \nonumber \\
&=& \bar{l}\Gamma_{lN}lH^0+H^0 \frac{1}{\delta_{ab}+1}\left(
\sum_{a \geq b}\bar{N}_a \left[ {\left( \Gamma_N \right)}
+{\left( \Gamma_N \right)}^C \right]_{ab} N_b \right)
\end{eqnarray}
where
\begin{eqnarray}
\Gamma_{(x)}&=&\left( P_LB_L^{(x)}+P_RB_R^{(x)} \right),\;
x=l,N,lN, \\
\bar{\Gamma}_l&=&\gamma_0  {\Gamma_l}^{\dagger}\gamma_0 \\
{\rm and} \hspace{3 cm} && \nonumber \\
\Gamma_N^C&=&C \Gamma_N^T C^{-1}.
\end{eqnarray}

Two remarks are necessary. Firstly, we would like to read out a form
of appropriate vertices for Majorana particles directly from Lagrangian
as is in the case of Dirac fermions both in QED and QCD. 
The above form of the Lagrangian satisfies this demand, for instance, the 
contents of
the rectangle brackets in Eqs.(C.2),(C.7) describe Majorana-Majorana-boson 
vertices (up to the $i$ factor).
Secondly, I consider two non-standard models in this Thesis. In the previous
Appendices I have written down the Lagrangian and/or couplings of 
particles in these models
I need through the Thesis, in agreement with the above denotations, so
vertices can be easily read from them.

Let's proceed to Feynman rules for any process involving Majorana particles.
These rules are based on the following procedure that we should perform:
\begin{itemize}
\item[(i)] attribute spinors to fermion lines on the given Feynman diagram (not
only to external lines);
\item[(ii)] set up a form of vertices;
\item[(iii)] build up a propagator in the way to get a Dirac type one;
\item[(iv)] establish related sign between different diagrams.
\end{itemize}

Ad.(i): Spinors \\

For the  Majorana-charged lepton coupling with charged gauge (Higgs) bosons
(let's call it `Majorana-Dirac' coupling) 
the spinors' attribution to the Majorana line depends on the nature of the
Dirac line \\
(a) For incoming Dirac particle (antiparticle) the outgoing Majorana fermion
must be treated as a particle (antiparticle) 

\input feynman.tex
\begin{picture}(5000,7000)
\thicklines
\drawline\fermion[\E\REG](12000,4800)[8000]
\drawline\fermion[\E\REG](12000,4600)[8000]
\put(500,4700){\scriptsize Dirac particle (antiparticle)}
\drawline\fermion[\E\REG](\fermionbackx,4700)[\fermionlengthx]
\put(29000,4700){\scriptsize $\bar{u}(v)$}
\drawline\photon[\S\REG](\pfrontx,\pfronty)[5]
\put(21000,2400){\scriptsize $W^{\pm}\;or\;H^{\pm}$}
\end{picture}
\vspace{0.75 cm}

(b) For outgoing Dirac particle (antiparticle) the incoming Majorana fermion
must be treated as a particle (antiparticle) 

\begin{picture}(4000,7000)
\thicklines
\drawline\fermion[\E\REG](8000,5700)[8000]
\drawline\fermion[\E\REG](\pbackx,5800)[8000]
\put(5000,\fermionfronty){\scriptsize $u ( \bar{v})$}
\drawline\fermion[\E\REG](\pfrontx,5600)[\fermionlengthx]
\put(25000,\pbacky){\scriptsize Dirac particle (antiparticle)}
\drawline\photon[\S\REG](\pfrontx,\pfronty)[5]
\put(17000,3400){\scriptsize $W^{\pm}\;or\;H^{\pm}$}
\end{picture}

We can see that for the Dirac-Majorana transition the attribution of spinors to 
Majorana lines is definitive. This is not the case for neutral 
Majorana-Majorana -neutral boson couplings (let's call it `Majorana-Majorana' couplings). 
Here we can treat a Majorana fermion as a particle or an antiparticle

\begin{picture}(4000,7000)
\thicklines
\global\seglength=1100
\global\gaplength=350
\drawline\fermion[\E\REG](8000,5700)[8000]
\put(5000,5500){\footnotesize $u ( \bar{v})$}
\drawline\fermion[\E\REG](\pbackx,5700)[8000]
\put(26000,5500){\footnotesize $\bar{u}(v)$}
\drawline\photon[\S\REG](\pfrontx,\pfronty)[5]
\put(17000,3400){\scriptsize $Z^0\;or\;H^0$}
\end{picture}

\setlength{\textwidth}{14 cm}
\setlength{\textheight}{19 cm}
\setlength{\topmargin}{1 mm}

If the `fermion flow' on the diagram
is opposite to the momentum flow then we can use the relation
$$u(\pm k)=v(\mp k)$$
to change its direction.

Ad.(ii) Vertices \\

(a) For a Dirac-Majorana coupling with a charged boson we have

\begin{picture}(4000,7000)
\thicklines
\global\seglength=1100
\global\gaplength=350
\drawline\fermion[\E\REG](2000,5800)[8000]
\drawline\fermion[\E\REG](2000,5600)[8000]
\drawline\fermion[\E\REG](\fermionbackx,5700)[\fermionlengthx]
\put(11000,3400){\scriptsize $W^{\pm}\;or\;H^{\pm}$}
\put(\pbackx,3000){\footnotesize $\left\{ \begin{array}{ll}
i\Gamma^{\mu}_l & \mbox{\rm for}\;\mbox{outgoing}\;W^-\;\mbox{or}\;
\mbox{incoming}\;W^+ \\
i\bar{\Gamma}^{\mu}_l & \mbox{for}\;\mbox{outgoing}\;W^+\;\mbox{or}\;
\mbox{incoming}\;W^-\\
i\Gamma_l & \mbox{for}\;\mbox{outgoing}\;H^-\;\mbox{or}\;\mbox{incoming}
\;H^+\\
i\bar{\Gamma}_l & \mbox{for}\;\mbox{outgoing}\;H^+\;\mbox{or}\;\mbox{incoming}
\;H^-
\end{array}
\right. $}
\drawline\photon[\S\REG](\pfrontx,\pfronty)[5]
\end{picture}

(b) For a Majorana-Majorana coupling with a neutral boson we have

\begin{picture}(4000,7000)
\thicklines
\global\seglength=1100
\global\gaplength=350
\drawline\fermion[\E\REG](2000,5700)[8000]
\put(2000,5000){\scriptsize $a$}
\drawline\fermion[\E\REG](\pbackx,5700)[8000]
\put(17000,5000){\scriptsize $b$}
\put(\pbackx,2000){\footnotesize $\left\{ \begin{array}{ll}
i\left[ \Gamma^{\mu}_{N}+{\Gamma^{\mu}_N}^C \right]_{ab} & \mbox{for}\;Z^0 \\
& \\
i\left[ \Gamma+\Gamma^C \right]_{ab} & \mbox{for}\;H^0  
\end{array}
\right. $}
\drawline\photon[\S\REG](\pfrontx,\pfronty)[5]
\put(11000,3000){\scriptsize $Z^0\;or\;H^0$}
\end{picture}

\setlength{\textwidth}{14 cm}
\setlength{\textheight}{19 cm}
\setlength{\topmargin}{1 mm}

\newpage

Ad.(iii) Propagators

We build up an amplitude in a way to get only one type of propagator for a
virtual Majorana particle, the same as for the Dirac case which we
build from u-type spinors

$$u(k)\bar{u}(k) \rightarrow i\sum\limits_{spin(\lambda)}\frac{u(\vec{k},
\lambda)\bar{u}(\vec{k},\lambda)}{k^2-m^2+i\epsilon}=\frac{i(\hat{k}+m)}
{k^2-m^2+i\epsilon} \equiv iS(k) $$

To do it we may sometimes need useful relations between spinors

\begin{equation}
\bar{v}_bOu_a=\bar{v}_aO^Cu_b;\;\;\;\bar{u}_bOu_a=\bar{u}_aO^Cu_b
\end{equation}
where $O=(\Gamma^{\mu},\Gamma)$ and $O^C=CO^TC^{-1}$.

If both (Majorana) spinors a and b describe external particles we have to take
relations (C.11) with the minus sign. \\

Ad.(iv): Sign convention 

The amplitude is used to calculate the cross section so we need to know only
a relative sign between various diagrams. This relative sign can be 
established as follows:
\begin{itemize}
\item choose any Feynman diagram which we call a reference diagram. 
In this diagram fermions appear in a given order;
\item compare all the other diagrams with the reference one. Permute fermions
in their `fermion chain' to get the same order as in the reference one;
\item if parity of the permutation is odd (even) then we change (unchange)
signature of the amplitude.
\end{itemize}
Examples can be found in \cite{pr1} (see also \cite{fey}).
These rules can be applied to loops diagrams, too (\cite{pr1},\cite{dub}). 


\newpage
\end{document}